\documentclass{aa}  

\usepackage{graphicx, caption}
\usepackage{natbib}
\usepackage{txfonts}
\usepackage{amsmath}
\usepackage{multirow}
\usepackage{float}
\usepackage[colorlinks=true,     linkcolor=blue, citecolor=blue, filecolor=blue, urlcolor=blue]{hyperref}
\usepackage{orcidlink}
\usepackage{comment}
\usepackage{setspace}
\begin{document}


   \title{Physical characterization of the FeLoBAL outflow in SDSS J0932+0840: Analysis of VLT/UVES observations }

   \author{Mayank Sharma
          \inst{1}\fnmsep\thanks{\email{mayanksh@vt.edu}} \orcidlink{0009-0001-5990-5790},
          Nahum Arav
          \inst{1}\orcidlink{0000-0003-2991-4618},
          Kirk T. Korista
          \inst{2}\orcidlink{0000-0003-0944-1008},
          Manuel Bautista
          \inst{2}\orcidlink{0000-0001-6837-3055},
          Maryam Dehghanian
          \inst{1}\orcidlink{0000-0002-0964-7500} ,\\
          Doyee Byun
          \inst{1, 3}\orcidlink{0000-0002-3687-6552},
          Gwen Walker
          \inst{1}\orcidlink{0000-0001-6421-2449},
          \and
          Sasha Mintz
          \inst{1}\orcidlink{0009-0009-0147-6485}
          }

   \institute{Department of Physics, Virginia Tech, Blacksburg, VA 24061, USA
         \and
             Department of Physics, Western Michigan University, 1120 Everett Tower, Kalamazoo, MI 49008-5252, USA
             \and
             Department of Astronomy, University of Michigan, Ann Arbor, MI, 48109, USA
             }

   \date{Received October 24, 2024; accepted December 9, 2024}

 
  \abstract
   {The study of quasar outflows is essential for understanding the connection between active galactic nuclei (AGN) and their host galaxies. We analyzed the VLT/UVES spectrum of quasar SDSS J0932+0840 and identified several narrow and broad outflow components in absorption, with multiple ionization species including \ion{Fe}{II}. This places it among the rare class of outflows known as iron low-ionization broad absorption line outflows (FeLoBALs).}
   {We studied one of the outflow components to determine its physical characteristics by determining the total hydrogen column density, the ionization parameter, and the hydrogen number density. Through these parameters, we obtained the distance of the outflow from the central source, its mass outflow rate, and its kinetic luminosity, and we constrained the contribution of the outflow to the AGN feedback.}
   {We obtained the ionic column densities from the absorption troughs in the spectrum and used photoionization modeling to extract the physical parameters of the outflow, including the total hydrogen column density and ionization parameter. The relative population of the observed excited states of \ion{Fe}{II} was used to model the hydrogen number density of the outflow.}
   {We used the \ion{Fe}{II} excited states to model the electron number density ($n_e$) and hydrogen number density ($n_H$) independently and obtained $n_e$ $\simeq$ $10^{3.4}$ cm$^{-3}$ and $n_H$ $\simeq$ $10^{4.8}$ cm$^{-3}$. Our analysis of the physical structure of the cloud shows that these two results are consistent with each other. This places the outflow system at a distance of $0.7_{-0.4}^{+0.9}$ kpc from the central source, with a mass flow rate ($\dot{M}$) of $43^{+65}_{-26}$ $M_\odot$ yr$^{-1}$ and a kinetic luminosity ($\dot{E_k}$) of $0.7^{+1.1}_{-0.4}$ $\times$ $10^{43}$ erg s$^{-1}$. This is $0.5^{+0.7}_{-0.3}$ $\times$ $10^{-4}$ of the Eddington luminosity ($L_{Edd}$) of the quasar, and we thus conclude that this outflow is not powerful enough to contribute significantly toward AGN feedback.}
  {}


    \titlerunning{FeLoBAL outflow in SDSS J0932+0840}
   \authorrunning{Mayank Sharma et al.}
   \maketitle
%

\section{Introduction}

    Quasar outflows are detected as blueshifted absorption troughs relative to the systemic redshift of the active galactic nuclei (AGN). They are observed ubiquitously and are invoked as the primary mechanism in the quasar mode of AGN feedback. They may play an important role in the evolution of the supermassive black hole \citep[SMBH; e.g.,][]{1998A&A...331L...1S,begelman2005self,hopkins2009small,liao2024rabbits}, in that of the host galaxy \citep[e.g.,][]{di2005energy,ostriker2010momentum,choi2017physics, cochrane2023impact}, in the enrichment of the intergalactic medium \citep[e.g.,][]{moll2007simulations,fabjan2010simulating,ciotti2022parameter}, and in cooling flows in clusters of galaxies \citep[e.g.,][]{2005RSPTA.363..655B,bruggen2009self,hopkins2016stellar,weinberger2023active}.
    
    Based on the width of the detected absorption lines (defined as continuous absorption with a normalized flux < 0.9), quasar outflows are commonly classified into one of the three following categories: Broad absorption lines (BALs) are defined as having a velocity width of $\Delta \textrm{v} \gtrsim$ 2000 km s$^{-1}$, and they are found in $\sim$ 20\% of the quasar spectra \citep{weymann1991comparisons, hewett2003frequency, reichard2003continuum, knigge2008intrinsic}, mini-BALs have velocity widths of 500 $\lesssim \Delta \textrm{v} \lesssim$ 2000 km s$^{-1}$ and an incidence rate of $\sim$ 10\% \citep{hamann2004asp,hidalgo2012variability,moravec2017hst,chen2021catalog}, and finally, narrow absorption lines (NALs) with a velocity width of $\Delta \textrm{v} \lesssim$ 500 km s$^{-1}$ are the most common features and are detected in $\sim$ 60\% of the quasar spectra \citep{vestergaard2003occurrence, ganguly2008fraction, stone2019narrow}.

    Broad absorption line quasar outflows (BALQSO) have further been classified into three categories based on the ionization state of the observed absorption lines in their spectra \citep{hall2002unusual}. High-ionization BALQSOs (HiBALs) only show absorption from high-ionization species, such as \ion{C}{iv}, \ion{Si}{iv}, \ion{N}{v}, and \ion{O}{vi}, and they are most common and observed in all BALQSOs. \citep{voit1993,dai2012intrinsic}. Low-ionization BALQSOs (LoBALs) show absorption lines from the high-ionization species, but are also accompanied by lower-ionization species such as \ion{Mg}{II}, \ion{C}{II}, and \ion{Al}{III}. They are observed in $\sim$ 10 \% of the BALQSOs \citep{weymann1991comparisons,trump2006catalog}. The rarest class are iron low-ionization BALQSOs (FeLoBALs), which show absorption from both high- and low-ionization species along with \ion{Fe}{II} or \ion{Fe}{III}. They make up just $\sim$ 0.3\% of the quasar population \citep{trump2006catalog}.

    Theoretically, these classifications for quasar outflows are understood through two different models. In the context of an orientation-based model, BALQSOs are observed as a result of the intersection of our line of sight with outflowing gas that is launched from a narrow range of radii on the accretion disk \citep{murray1995accretion,elvis2000structure}. The different ionization subclasses then relate to the column density of the material they encounter, with HiBALS having the smallest column and FeLoBALs having the largest, extending beyond the \ion{H}{II} region \citep{arav2001hst, korista2008physical, lucy2014tracing}. The opening angle for this line of sight would be small \citep{green2001chandra, gallagher2006exploratory, morabito2011suzaku} and thus explains the rarity of FeLoBALs. An alternative evolutionary model suggests that FeLoBALs might represent a short-lived phase in which a young quasar is observed in the process of breaking through a cocoon of enveloping dust and gas \citep{voit1993}. 

     The FeLoBALs usually show a plethora of absorption troughs from ground-state and various excited-state transitions, which makes them an excellent diagnostic of the physical state of the outflowing gas. \citet{arav2001intrinsic} and \cite{de2001keck, de2002keck, de2002intrinsic} presented some of the first detailed photoionization analyses of FeLoBAL quasars based on Keck High Resolution Echelle Spectrometer (HIRES) spectra and obtained constraints on the temperature, the ionization parameter ($U_H$), the electron number density ($n_e$), and the distance from the emission source ($R$) for the outflows. \cite{korista2008physical}, \citet{arav2008measuring}, \citet{moe2009quasar}, \citet{dunn2010quasar}, and \citet{bautista2010distance} used high-resolution spectra with a high signal-to-noise ratio from the Very Large Telescope (VLT) and the Astrophysical Research Consortium telescope (ARC). This allowed them to improve the accuracy of the column density and number density measurements and obtain mass flux and kinetic luminosity for some outflows, thus constraining the extent of their contribution to AGN feedback processes. \cite{leighly2018z} introduced the novel spectral synthesis procedure \textit{SimBAL}, which has led to an impressive increase in the number of FeLoBALs that were subjected to a detailed spectral analysis. \cite{choi2022physical} presented a \textit{SimBAL} analysis of 50 low-redshift FeLoBAL quasars that covered a wide range in the parameter space for $U_H$, $n_e$, and $R$, and introduced a new class of loitering FeLoBAL outflows. \cite{choi2022physical3} and \cite{leighly2022physical} extended this analysis further to the geometry of the outflows and the optical line emission properties for the host quasars. Other recent studies of individual quasar outflows have led to the discovery of increasingly interesting properties and phenomenon: \cite{choi2020discovery} discovered a remarkably powerful FeLoBAL outflow with the highest observed velocity and kinetic luminosity so far, \cite{xu2021physical} found differences in the physical conditions for the \ion{Fe}{ii} and \ion{Fe}{iii} formation in Q0059–2735, \cite{byun2022farthest} found the farthest known mini-BAL outflow system at $\sim$ 67 kpc and along with  \cite{walker2022high}, an extremely high mass outflow rate that might contribute significantly to AGN feedback processes.  

\begin{figure}
         \centering
         \includegraphics[width=1\linewidth]{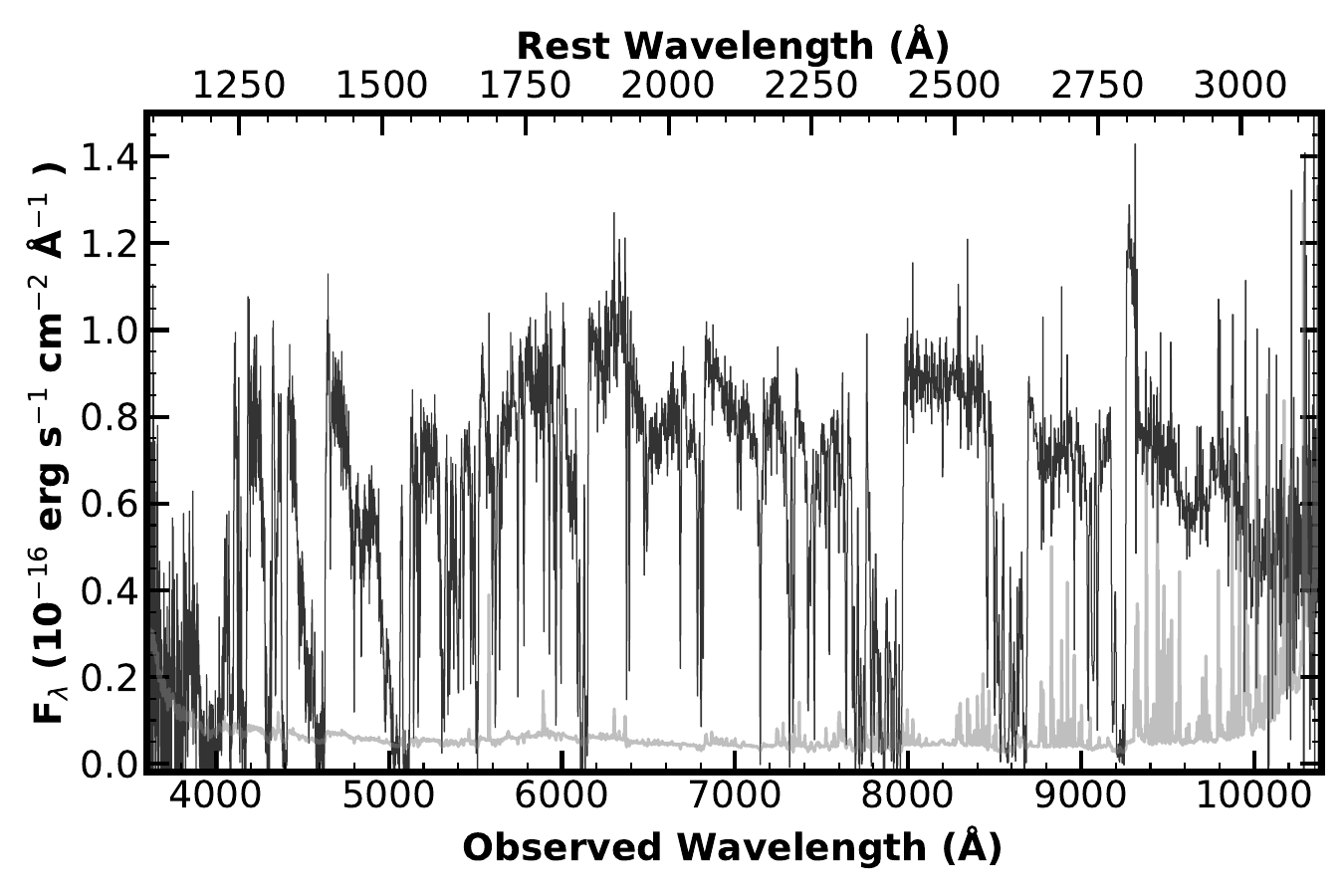}
         \caption{Non-normalized SDSS spectra of J0932+0840 (2012 epoch). The flux density ($F_{\lambda}$) is shown in black, and the error is plotted in gray.}
         \label{fig:SDSS}
\end{figure}

\begin{figure}
         \centering
         \includegraphics[width=1\linewidth]{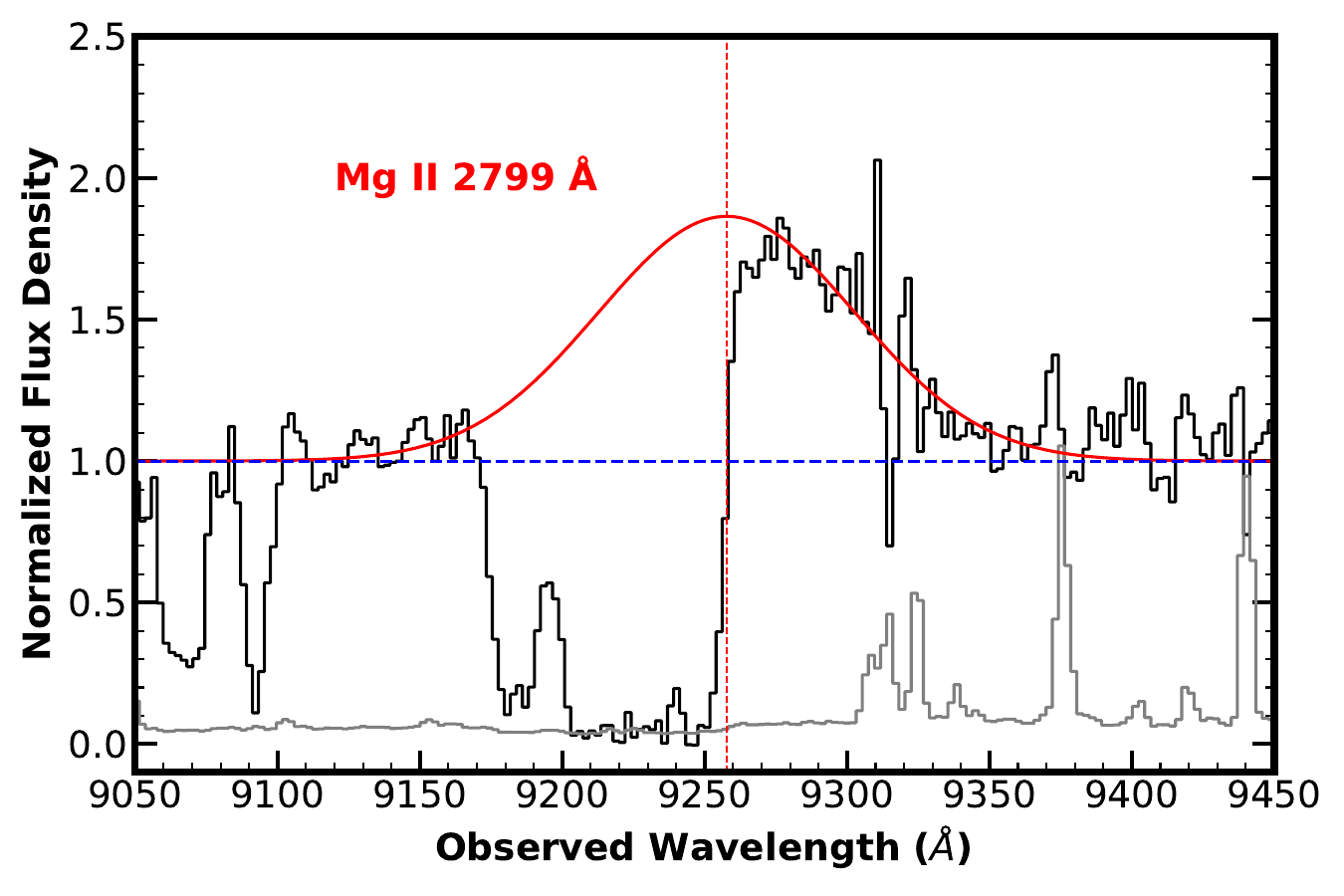}
         \caption{Gaussian model fit for the \ion{Mg}{ii} 2799 emission feature (shown in red). Due to high absorption, the data points for the fit were selected manually. The dashed red line marks the centroid of the best-fit model we used to determine the redshift of the quasar. The dashed blue line shows the continuum level. The error on the flux is shown in gray.}
         \label{fig:MgII}
\end{figure}

\begin{figure*}
         \centering
         \includegraphics[width=0.95\linewidth]{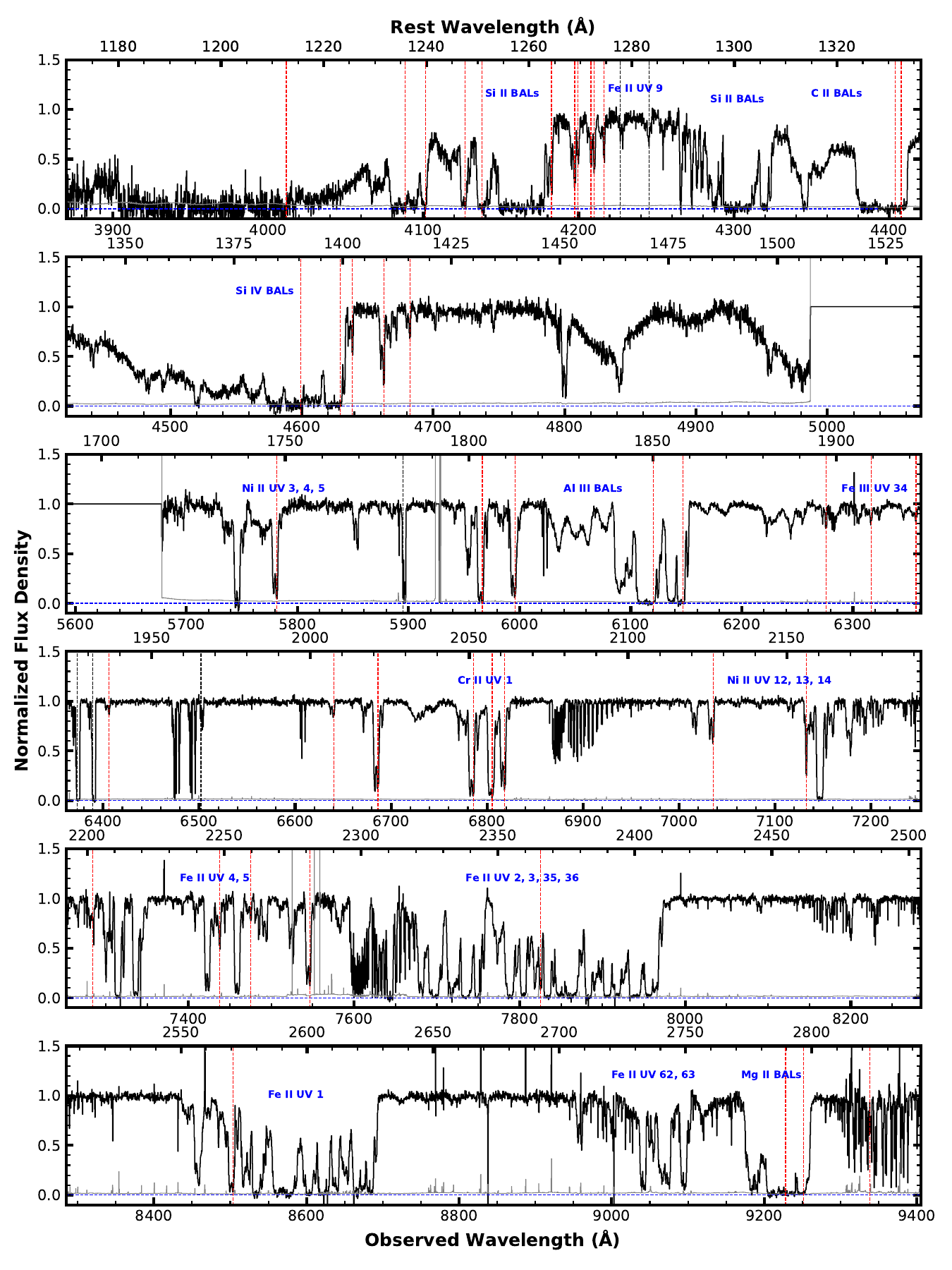}
         \caption{Normalized UVES spectrum of J0932-0840. Important features and multiplets are marked in blue, and the identified absorption troughs from the primary outflow system at $v\approx-720$ km s$^{-1}$ are marked with dashed red lines. The zero flux level is shown by the dashed blue lines. The dashed black lines mark troughs from an intervening system at $z$ = 1.28.}
         \label{fig:spectrum}
\end{figure*}

We present a study of the Ultraviolet and Visual Echelle Spectrograph (UVES) spectrum of the quasar SDSS J093224.48+084008.0 (hereafter, J0932+0840), obtained with the VLT. Section \ref{data} describes the observation and how we obtained the data for our analysis. In Sect. \ref{spec_analysis} we analyze the spectrum to obtain the column densities of the observed troughs, along with the photoionization modeling of the outflow. Section \ref{sec:den} discusses our diagnostics for the outflow electron and hydrogen number densities based on the population of the observed excited states. Section \ref{energy} presents the energetics of the outflow. Section \ref{discuss} discusses the observed variability between the different epochs, along with a reevaluation of properties from a few previous FeLoBAL studies. Finally, we summarize and conclude the paper in Sect. \ref{summary}. Throughout the paper, we adopt a standard $\Lambda$CDM cosmology with h = 0.677, $\Omega_m$ = 0.310 and $\Omega_{\Lambda}$ = 0.690 \citep{planck2018}.

\section{Data overview} \label{data}

\subsection{Observation and data acquisition} 

J0932+0840 (J2000: RA=09:32:24.48; DEC: +08:40:08.0) was observed with VLT/UVES on 30 April 2008 with a total exposure time of 28.1 ks as part of program 081.B-0285 (PI: Benn). The data cover a spectral range of 3758-10250 $\AA$, with a resolution of R$\approx$ 48,000 and an average signal-to-noise ratio S/N $\approx$ 30 per resolution element. The data were then reduced and normalized by \cite{murphy2019uves} as part of the first data release (DR1) of the UVES Spectral Quasar Absorption Database (SQUAD). Figure \ref{fig:spectrum} shows the obtained normalized data. 

J0932+0840 was also observed on two occasions (20 December 2003 and 26 January 2012) as part of the Sloan Digital Sky Survey (SDSS). While our primary analysis is based on the SQUAD data with their higher resolution and S/N, it lacks information about any emission features as it is normalized. We therefore used the SDSS spectrum (Fig. \ref{fig:SDSS}) to (i) obtain the redshfit of the quasar (Sect. \ref{sec:redshift}), (ii) scale the flux of our spectral energy distribution (SED) to obtain the bolometric luminosity (Sect. \ref{sec:distance}), and (iii) obtain the black hole mass (Sect. \ref{sec:bhmass}). 

\subsection{Redshift determination} \label{sec:redshift}

The second epoch of the SDSS observation (Fig. \ref{fig:SDSS}) covers the \ion{Mg}{II} $\lambda \lambda$ 2796.35, 2803.53   emission feature. This feature can be used to obtain the redshfit of the quasar. We first modeled the underlying continuum emission with a power law and then modeled the \ion{Mg}{II} emission feature with a single Gaussian. The result of this fit is shown in Fig. \ref{fig:MgII}. This led to a redshift of $z$ = 2.308 $\pm$ 0.001. This is close to the redshift estimate obtained by \cite{yi2019variability} ($z$ = 2.300) in their study of 134 Mg II broad absorption line (BAL) quasars including J0932+0840. We note, however, that these redshift values disagree with the value determined by \cite{hewett2003frequency} ($z$ = 2.341231 $\pm$ 0.001884). These authors determined this value by modeling the 1900 blend, which includes \ion{Al}{iii} 1857 ($\lambda \lambda$ 1854.72, 1862.78 ), \ion{Si}{iii]} 1892 ($\lambda$ 1892.03 ), and the \ion{C}{III]} 1909 ($\lambda$ 1908.73 ) emission lines in the first epoch of SDSS observation, as it does not cover the \ion{Mg}{II} emission feature. As this blend includes multiple emission features of similar strengths, it is more prone to uncertainty than the fit obtained for \ion{Mg}{II}, and we therefore used $z$ = 2.308 for our analysis.

\section{Spectral analysis} \label{spec_analysis}

\subsection{Identifying outflow systems} \label{colden}

The UVES spectrum of J0932+0840 shows a large number of absorption systems ranging in velocities from $-$500 to $-$5000 km s$^{-1}$. The lowest-velocity systems ($\sim$ $-$500, $-$700 and $-$850 km s$^{-1}$) are narrow absorption lines ($\Delta v \sim$ 100 km s$^{-1}$) and are observed in most species, including low- and high-ionization lines. The higher-velocity systems ($\sim$ $-$1300, $-$4200 and $-$5000 km s$^{-1}$) are much broader (with $\Delta v \gtrsim$ 2000 km s$^{-1}$, and thus classified as BALs), and are only seen in a few species such as \ion{Si}{ii}, \ion{C}{ii}, \ion{Mg}{ii}, \ion{Al}{iii},  \ion{Si}{iv}, and \ion{C}{iv} (the latter is only covered in the SDSS spectrum). 
The focus of this paper is on the system with velocity $\sim -700$ km s$^{-1}$ (hereafter S2) as it shows absorption from the most species and is thus rich in diagnostics.

\subsection{Column density determination} \label{sec:coldendet}

In order to constrain the physical characteristics of an outflow system, it is important to obtain the column densities of the observed ionic species. For a given ionic transition with a rest wavelength $\lambda$ (in Angstroms) and oscillator strength $f$, the column density $N_{ion}$ is obtained using \citep{savage1991analysis}
\begin{equation}{\label{nion}}
    N_{ion} = \frac{m_e c}{\pi e^2 f \lambda} \int \tau(v) \textrm{ dv} = \frac{3.8 \times 10^{14}}{f \lambda} \int \tau(v) \textrm{ dv} \textrm{\hspace{0.3cm}[cm$^{-2}$]},
\end{equation}
where $m_e$ is the electron mass, $c$ is the speed of light, $e$ is the elementary charge, and $\tau(v)$ is the optical depth profile for the transition, which is based on the choice of the absorber model. The simplest model, known as the apparent optical depth (AOD) model, assumes a homogeneous outflow that completely covers the source. In this case, the normalized intensity profile $I(v)$ is then simply related to $\tau(v)$ as
\begin{equation}
    I(v) = e^{-\tau(v)}.
\end{equation}
This model does not account for partial line-of-sight (LOS) covering and saturation effects in line centers and can thus underestimate the column densities. For saturated troughs, only lower limits to the column densities can therefore be obtained with the AOD model. In the case of J0932+0840, however, we detected several weak transitions that have low oscillator strength values. These troughs are much shallower than the other deeper troughs from the same ion and can thus be considered unsaturated. We therefore regarded the derived column densities for these transitions as actual measurements. \par
When two or more lines from the same lower-energy level were detected for an ion, we employed a more physical inhomogeneous absorber model in which the gas distribution is approximated with a power law \citep{de2002effects, arav2005x}. The optical depth profile within the outflow is then described as 
\begin{equation}
    \tau_v(x) = \tau_{max}(v)x^a,
\end{equation}
where $x$ is the spatial dimension across our LOS, and $a$ is the power law index. Then, given two or more intensity profiles, we solved for $\tau_{max}$ and $a$ simultaneously. 

\begin{figure*}
         \centering
         \includegraphics[width=1\linewidth]{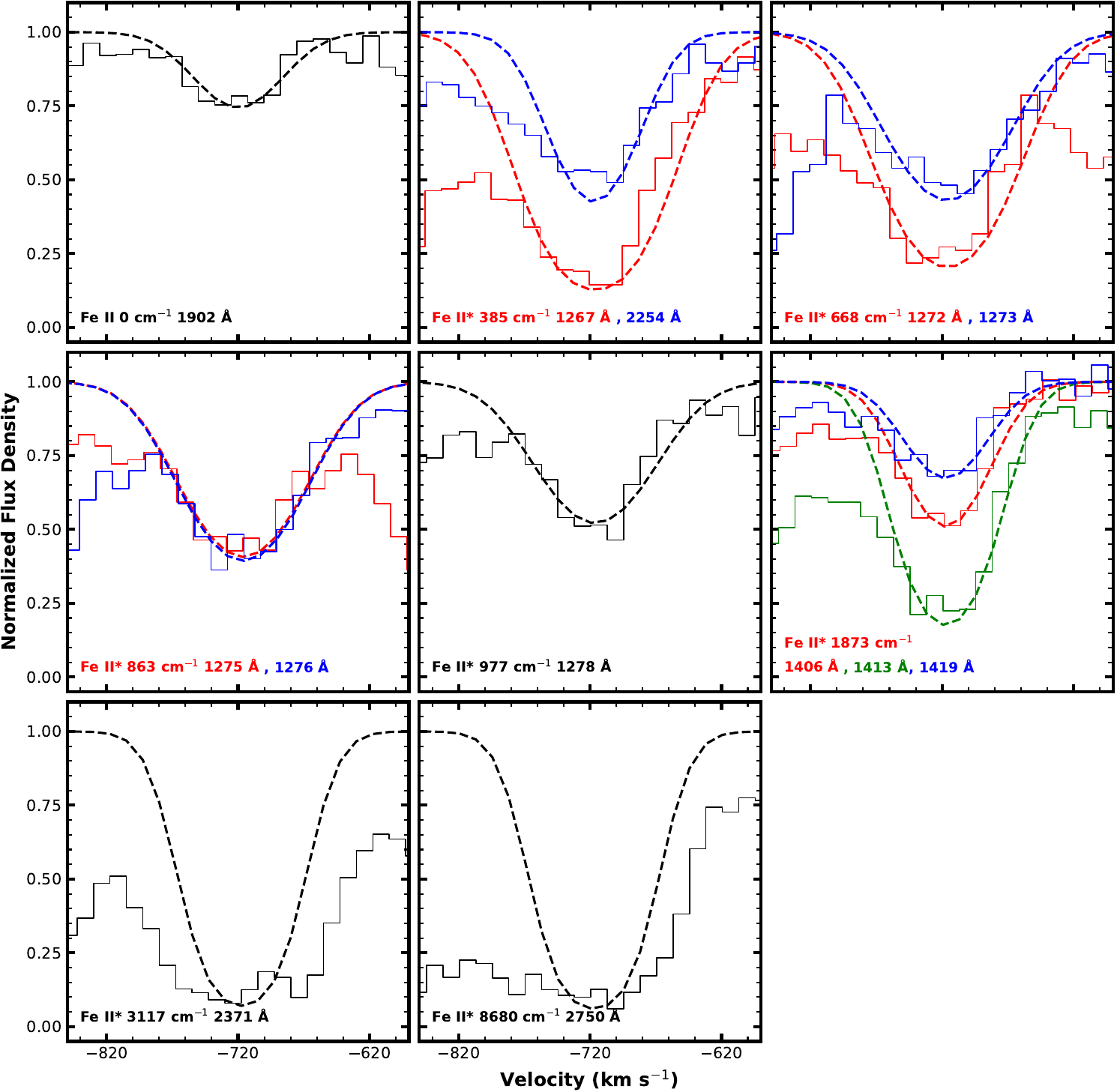}
         \caption{Identified \ion{Fe}{ii} troughs for outflow system S2. The histograms show the VLT/UVES spectra in velocity space, and the corresponding dashed curves show the modeled Gaussian for the absorption. The model is based on the 1278 $\AA$ transition, and its template with fixed centroid and width was scaled to match the depth of all other troughs. The lower-energy level of the transition and the wavelengths are given for each panel.}
         \label{fig:feiimodel}
\end{figure*}

The unsaturated absorption troughs of \ion{Fe}{ii} cover a velocity range -820 $\lesssim$ v $\lesssim$ -600 km s$^{-1}$ and were modeled by a Gaussian profile based on the \ion{Fe}{ii}* 1278 $\AA$ trough. The corresponding best-fit profile was scaled to match the other \ion{Fe}{II} troughs, and the final fits are shown in Fig. \ref{fig:feiimodel}. Except for the 7955 and 8680 $\textrm{cm}^{-1}$ levels, all other troughs appear to be unsaturated, and therefore, their AOD column densities were taken as measurements. We also detected multiple unsaturated transitions from the 385, 668, 863, and 1873 $\textrm{cm}^{-1}$ levels and therefore applied the power-law (PL) model to obtain their column densities. We find that for these transitions, the $N_{ion}$ values derived from the PL method are similar to the AOD model and therefore lend credibility to the AOD column densities derived for other species. We repeated this analysis for the other observed troughs in the spectrum and report their column densities in Table \ref{table:colden}. The reported error bars for the column density measurements include a 20\% relative error added in quadrature to account for systemic uncertainties, following the method described by \cite{xu2018mini}.  

\begin{table*}
\setlength{\tabcolsep}{10pt}
\renewcommand{\arraystretch}{1.33}
\caption{Absorption lines identified in the VLT spectrum of SDSS J0932+0840 and their measured column densities.}
\centering
\begin{tabular}{lcccc}
\hline\hline
Ion &$\lambda^a$&log($f$)$^b$&$E_{low}^c$ &$N_{ion}^d$ \\
    &  (\AA)    &        & (cm$^{-1})$&(log(cm$^{-2}$))\\  
\hline
\ion{H}{I}       &1215.67 & -0.38 & 0 & $> 14.57_{-0.23}$\\
\ion{He}{I}*     &2829.91 & -2.22 & 159856 & $14.14_{-0.11}^{+0.09}$\\
\ion{C}{II}      &1334.53 & -0.89 & 0 & $> 15.20_{-0.18}$\\
\ion{C}{II}*     &1335.71 & -0.94 & 63 & $> 15.24_{-0.32}$\\
\ion{N}{V}     &1238.82, 1242.80 & -0.80, -1.10 & 0 &$> 15.31_{-0.33}$\\
\ion{Mg}{II}     &2796.35, 2803.53 & -0.22, -0.52 & 0 &$> 14.52_{-0.29}$\\
\ion{Al}{III}     &1854.72, 1862.79 & -0.25, -0.55 & 0 &$> 14.70_{-0.16}$ \\
\ion{Si}{II}      &1808.01 & -2.60 &0&$16.57_{-0.15}^{+0.11}$\\
\ion{Si}{II}*     &1816.93 & -2.71 &287&$16.55_{-0.11}^{+0.09}$\\
\ion{Si}{IV}     &1393.75, 1402.77 & -0.29, -0.59 & 0 &$> 14.84_{-0.32}$\\
\ion{S}{II}     &1250.58, 1253.81 & -0.22, -1.92 & 0 &$> 16.44_{-0.31}$\\
\ion{Cr}{II}     &2056.25, 2062.24, 2066.16 & -0.99, -1.12, -1.29 & 0 &$> 15.08_{-0.10}$\\
\ion{Cr}{II}*     &2161.38 & -1.84* &42898&$14.95_{-0.11}^{+0.09}$\\
\ion{Mn}{II}      &2576.88 & -0.45 &0&$> 14.04_{-0.11}$\\
\ion{Fe}{II}      &1901.78 & $-4.10^{+0.30}_{-0.30}$ & 0 (0) & $16.77_{-0.36}^{+0.36}$\\
\ion{Fe}{II}*     &1267.42, 2253.82 & -1.85, -2.47 & 385 (1) & $15.50_{-0.10}^{+0.08}$\\
\ion{Fe}{II}*     &1271.98, 1272.61 & -1.85, -2.36  & 668 (2) & $15.77_{-0.10}^{+0.08}$\\
\ion{Fe}{II}*     &1275.14, 1275.78 & -2.15, -2.00 & 863 (3) & $15.59_{-0.10}^{+0.08}$\\
\ion{Fe}{II}*      &1277.66 & -1.85 & 977 (4) & $15.15_{-0.10}^{+0.08}$\\
\ion{Fe}{II}*     &1405.60, 1412.83, 1418.85 & -2.92, -2.47, -3.04* & 1873 (5) & $16.06_{-0.11}^{+0.09}$\\
\ion{Fe}{II}*      &2371.22 & -1.84 & 3117 (8) & $> 15.33_{-0.11}$\\
\ion{Fe}{II}*      &2749.99 & -0.86 & 8680 (11) & $> 14.32_{-0.10}$\\
\ion{Fe}{III}*     &1895.48, 1914.07, 1926.32 & -0.33, -0.49, -0.60 & 30089 & $13.18_{-0.11}^{+0.13}$\\
\ion{Co}{II}       &1941.28, 2012.17 & -1.47, -1.43 &0&$13.94_{-0.10}^{+0.08}$\\
\ion{Co}{II}*      &2286.86 & -0.51 &3350&$14.05_{-0.10}^{+0.08}$\\
\ion{Ni}{II}      &1751.91 & -1.54 & 0 &$>15.35_{-0.21}$ \\
\ion{Ni}{II}*     &2131.93 & -2.58* & 8394  &$15.53_{-0.11}^{+0.09}$ \\
\ion{Ni}{II}*     &2207.41, 2265.16 & -0.79, -0.83 &10115 & $13.59_{-0.17}^{+0.12}$\\
\ion{Zn}{II}      &2026.14 & -0.30 & 0 &$13.87_{-0.10}^{+0.08}$\\
\hline
\end{tabular}
\tablefoot{(a) Rest wavelength of the transition in vacuum. Multiple values, wherever mentioned, correspond to detection of multiple troughs for the given energy level of the ionic species. (b) Logarithm of the oscillator strength of the transition. These were taken from the National Institute of Standards and Technology (NIST) Atomic Spectra Database vers. 5.11, unless marked otherwise. The values marked with an asterisk were obtained from \cite{kurucz1995atomic}, and the value for the \ion{Fe}{II} 1902 transition was determined from our calculations described in Sect. \ref{sec:osc}. (c) Lower-energy level of the transition. For \ion{Fe}{ii}, the indexes used for the energy levels in the \textsc{Cloudy} modeling are shown in parentheses. (d) Column density of the ionic transition. Measurements were obtained for unsaturated troughs, whereas saturated troughs only lead to lower limits on the column density of the transition.}
\label{table:colden}
\end{table*}

\subsection{Calculating the oscillator strength for \ion{Fe}{II} transitions} \label{sec:osc}

As mentioned above, a special feature of the VLT spectrum of J0932+0840 is the detection of unsaturated troughs that correspond to transitions with very weak oscillator strengths. The \ion{Fe}{ii} 1901.78 ground-state transition is the weakest of these transitions, with a theoretically determined oscillator strength of 6.25 $\times$ 10$^{-5}$ \citep{kurucz1995atomic}, and is thus rarely observed. The only other detection in a quasar spectrum was reported by \cite{byun2024extreme}. This rarity means, however, that the oscillator strength value for the transition remains poorly constrained by observations and is prone to uncertainties. As this transition is of central importance in our work (see Sect. \ref{sec:den}), we performed our own calculations to obtain an estimate for the oscillator strength along with its uncertainties. 

We modeled the \ion{Fe}{ii} atomic system and computed gf values for dipole bound-bound transitions using the computer code \textsc{Autostructure} \citep{badnell2011breit}. \textsc{Autostructure} solves the Breit–Pauli Schrödinger equation with a scaled Thomas–Fermi–Dirac-Amaldi (TFDA) central-field potential to generate orthonormal atomic orbitals. Configuration interaction (CI) atomic state functions are built using these orbitals. The scaling factors of the potential for each orbital are generally optimized in a multiconfiguration variational procedure minimizing a sum over LS term energies or a weighted average of LS term energies. This optimization can be made on nonrelativistic LS calculations or with one-body relativistic operators.

Two types of corrections are enabled by the code after optimizing the orbitals. Perturbative term energy correction (TEC) can be applied to the multi-electronic Hamiltonian, which adjusts theoretical LS term energies closer the centers of gravity of the available experimental multiplets. Level energy corrections (LEC) can also be applied to shift the theoretical energy levels to reproduce matching level energy separations when computing f values and transition probabilities.

For large calculations such as ours, \textsc{Autostructure} was run from our own \textsc{Python} wrapper for the semi-automatic generation of input files with large configuration expansions, analysis of output files, comparisons of energies and radiative data, and TEC and LEC corrections. We carried out numerous calculations for the \ion{Fe}{ii} system, systematically added configurations to the CI expansion, and fully optimized the orbitals with every expansion. The aim was to include within the computational limits all configurations that contribute to improving the energies of the terms belonging to the 3d$^6$4s, 3d$^7$, 3d$^5$4s$^2$, 3d$^6$4p, and 3d$^5$4s4p configurations relative to the experimental values, as well as the convergence of the gf values for the transitions of interest. The energies of the 3d$^7$~a~$^4$F levels relative to levels the 3d$^6$4s~a~$^6$D ground term are particulaarly challenging and important because all observed transitions in this work arise from them. 

Our final expansion included 11 orbitals (1s, 2s, 2p 3s, 3p, 3d, 4s, 4p, 4d, 5s, 5p) in 61 configurations with 1s and 2s closed-core orbitals: 3d$^7$, 3d$^6nl~$($n$ = 4 – 5, $l$ = 0 – 2),  3d$^5nln’l’~$($n$ = 4 – 5, $l$ = 0 – 2, $n’$ = 4 – 5, $l’$ = 0 – 2), 3p$^5$3d$^8$, 3p$^5$3d$^7nl~$($n$ = 4 – 5, $l$ = 0 – 2),  3p$^5$3d$^64snl~$($n$ = 4 – 5, $l$ = 0 – 2), 3p$^4$3d$^9$, 3p$^4$3d$^8nl~$($n$ = 4 – 5, $l$ = 0 – 2),  3p$^4$3d$^7nl^2~$($n$ = 4 – 5, $l$ = 0 – 2), 3s3p$^6$3d$^8$, 3s3p$^6$3d$^7nl~$($n$ = 4 – 5, $l$ = 0 – 2),  3s3p$^6$3d$^6$4s$^2$, 2p$^5$3s$^2$3p$^6$3d$^8$, 2p$^5$3s$^2$3p$^6$3d$^7nl~$($n$ = 4 – 5, $l$ = 0 – 2),  2p$^5$3s$^2$3p$^6$3d$^6$4s$^2$,
2p$^6$3p$^6$3d$^9$, 2p$^6$3p$^6$3d$^8$4s, 2p$^6$3p$^6$3d$^7$4s$^2$,
2p$^6$3p$^6$3d$^7$4s4p, 2p$^4$3s$^2$3p$^6$3d$^9$, 2p$^4$3s$^2$3p$^6$3d$^8$4s,  2p$^4$3s$^2$3p$^6$3d$^7$4l$^2~$($l$ = 0 – 2).

An important issue to highlight with these calculations is that when an acceptable level of converse was achieved in terms of the predicted energies, a further slight optimization of the 3d and 4s orbitals had to be performed manually on fine-structure energy levels rather than the numerical optimization based on LS terms incorporated in \textsc{Autostructure} for the correct ordering and energy separations between the 3d$^7$~a~$^4$F and 3d$^6$4s~a~$^6$D levels. The energy separation between these is comparable to the magnitude of the two-body relativistic corrections, hence optimizations in LS-coupling of these terms is unphysical and becomes severely disrupted in final jj-coupling computations. Moreover, the relativistic jj-coupling correction on level energies is highly nonlineal upon initial LS term energies. Because of the magnitude of two-body energy corrections on levels of the 3d$^7$~a~$^4$F and 3d$^6$4s~a~$^6$D, no perturbative TEC can be employed. The orbitals must instead be optimized by hand to obtain accurate level energies, and simple LEC was then applied for the final computation of gf values.

Our calculation resulted into 669 energy levels from the five configurations listed above. These levels yield nearly 40,000 dipole-allowed transitions and over 100,000 dipole-forbidden transitions. This extensive dataset enables the proper identification of the troughs observed in our experiment, many of which were not found in previously published tabulations of \ion{Fe}{ii} gf values. We recall that the \ion{Fe}{ii} troughs identified in the VLT/X-Shooter spectrum of J0932+0840 are intrinsically weak transitions, and their $gf$ values therefore carry a significant uncertainty. The radiative rates found for these transitions in the NIST database have accuracy ratings of D or D+ (i.e., an uncertainty $\gtrsim$ 50 \%). Table \ref{FeII_gfvals} compares the $gf$  values from the NIST database with our calculations for a sample of transitions.

For the transitions with an available $gf$ value in the NIST database, we used this value in our analysis. For the \ion{Fe}{II} 1901.78 \r{A} transition, we used the value obtained from our calculations, with log($gf$) = $-3.1^{+0.3}_{-0.3}$. This agrees with the value reported in \cite{kurucz1995atomic}. Our calculations importantly also gave us an estimate for the associated uncertainty, which we propagated in our calculation for the \ion{Fe}{II} ground-state column density (Table \ref{table:colden}) and thus in the remaining analysis.   

\begin{table*}
\setlength{\tabcolsep}{10pt}
\renewcommand{\arraystretch}{1.33}
\caption{Comparison between the $gf$ values obtained from our calculations and the values available in the NIST database for a sample of transitions.}
\centering
\begin{tabular}{lcccc}
\hline\hline

Transition & Wavelength  & \multicolumn{2}{c}{log($gf$)} \cr
 & $(\AA)$ & NIST & This work \cr
 \hline
3d$^6$4s~a~$^6$D$_{7/2}$         -- 3d$^5$4s4p~x~$^6$P$^o_{5/2}$        &1267.43&       -0.95&  -0.70   \cr             
3d$^6$4s~a~$^6$D$_{5/2}$         -- 3d$^5$4s4p~y~$^6$P$^o_{5/2}$        &1271.99&       -1.08&  -0.82   \cr             
3d$^6$4s~a~$^6$D$_{5/2}$         -- 3d$^5$4s4p~y~$^6$P$^o_{3/2}$        &1272.62&       -1.58&  -1.11   \cr             
3d$^6$4s~a~$^6$D$_{3/2}$         -- 3d$^5$4s4p~y~$^6$P$^o_{5/2}$        &1275.15&       -1.60&  -1.29   \cr             
3d$^6$4s~a~$^6$D$_{3/2}$         -- 3d$^5$4s4p~y~$^6$P$^o_{3/2}$        &1275.78&       -1.40&  -0.92   \cr             
3d$^6$4s~a~$^6$D$_{1/2}$         -- 3d$^5$4s4p~y~$^6$P$^o_{3/2}$        &1277.65&       -1.54&  -1.06   \cr             
3d$^7$~a~$^4$F$_{9/2}$  -- 3d$^6$4p~z~$^2$F$^o_{7/2}$   &1405.61&       -1.90&  -2.32   \cr             
3d$^7$~a~$^4$F$_{9/2}$  -- 3d$^6$4p~z~$^4$D$^o_{7/2}$   &1412.84&       -1.47&  -1.30   \cr

 \hline
\end{tabular}
\label{FeII_gfvals}
\end{table*}

\subsection{Photoionization modeling}\label{nvu}

Photoionization dominates the ionization equilibrium in quasar outflows. Therefore, an outflow is characterized by its total hydrogen column density ($N_H$) and ionization parameter ($U_H$), which is related to the rate of ionizing photons emitted by the source ($Q_H$) by
\begin{equation} \label{uh}
    U_H = \frac{Q_H}{4\pi R^2 n_H c},
\end{equation} 
where $R$ is the distance between the emission source and the observed outflow component, $c$ is the speed of light, and $n_H$ is the hydrogen number density. $Q_H$ is determined by the choice of the SED that is incident upon the outflow and the redshift of the object. 

\begin{figure}
   \centering
   \resizebox{\hsize}{!}
        {\includegraphics[width=1.20\linewidth]{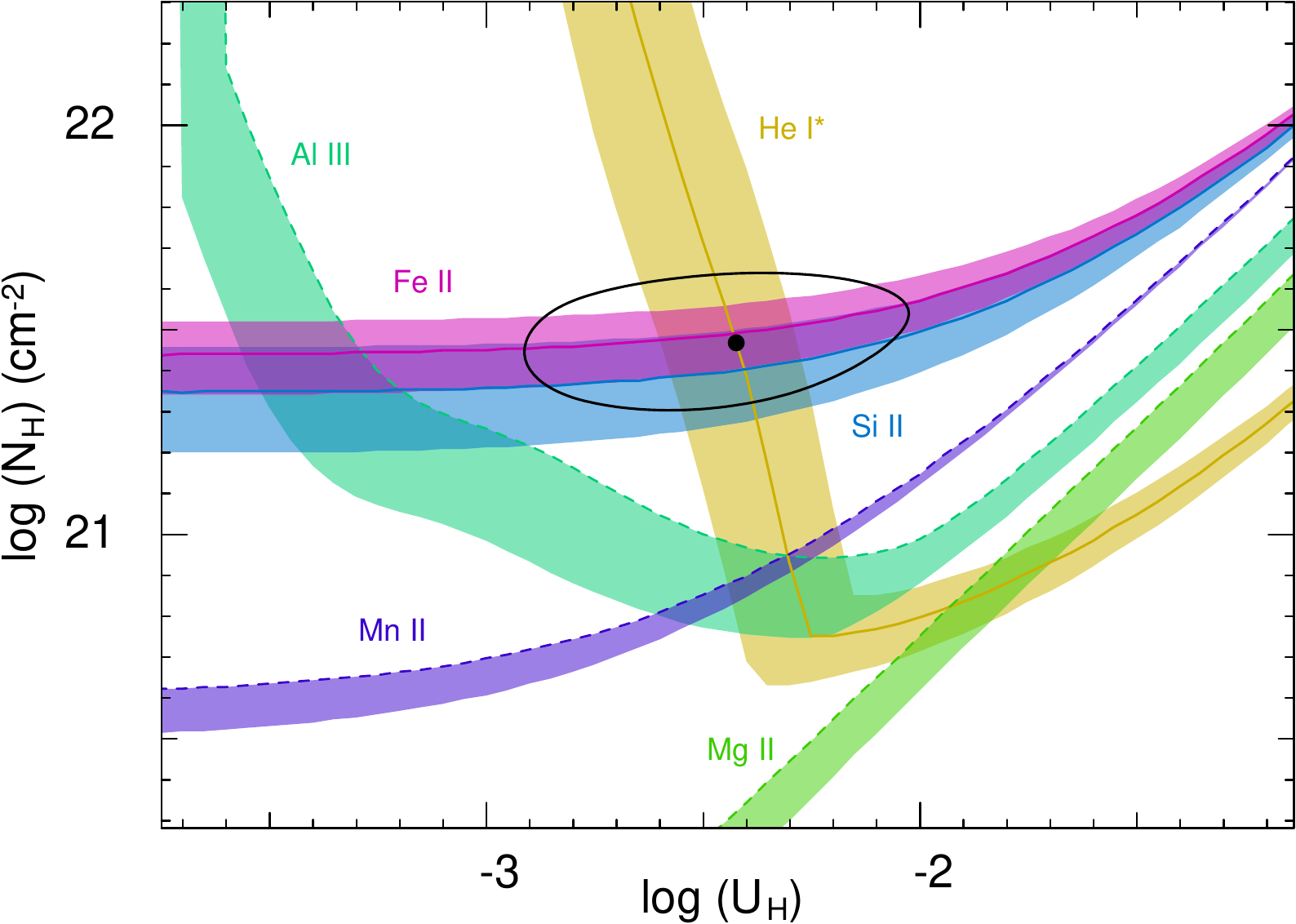}}
      \caption{ $\log{N_H}$ vs. $\log{U_H}$ phase-space plot, with constraints based on the measured total ionic column densities (sum of the column densities of the ground state and all the observed excited states). The measurements are shown as solid curves, and the dashed curves show lower limits. The shaded regions denote the errors associated with each $N_{ion}$. The phase-space solution with minimized $\chi^2$ is shown as a black dot surrounded by a black eclipse indicating the 1$\sigma$ error.}
         \label{fig:nvu}
   \end{figure}

Using the obtained column densities, we constrained $N_H$ and $U_H$ and thus determined the physical state of the gas. This was done using version C23.01 of the spectral synthesis code \textsc{Cloudy} \citep{cloudy2023}, which solves the ionization equilibrium equations. The outflow was modeled as a plane-parallel slab with a constant $n_H$ (= $10^5$ cm$^{-3}$; see Sect. \ref{sec:hden}) and solar abundance, and it was irradiated upon by the modeled SED from quasar HE0238-1904 \citep{arav2013quasar}, which is the best empirically determined SED in the extreme UV where most of the ionizing photons come from. Using the method described by \cite{arav2001hst} and \cite{edmonds2011galactic}, we varied $N_H$ and $U_H$ in small steps and kept all other parameters constant. This led to a grid of models in the ($N_H$, $U_H$) parameter space, where each point in the grid predicted the total column densities (i.e., the sum of the column densities of the ground state and all the excited states) of all relevant ions. We then mapped the measured ionic column densities ($N_{ion}$) against the model predictions, which resulted in a phase-space plot containing the ionic equilibrium solutions for all observed ions, as shown in Fig. \ref{fig:nvu}. In this space, an ($N_H$, $U_H$) value that is within one standard deviation of all the solutions would provide a viable model. The best-fit solution determined by minimizing $\chi^2$ led to log $N_H$ = $21.47^{+0.17}_{-0.16}$ [cm$^{-2}$] and log $U_H$ = $-2.4^{+0.4}_{-0.5}$. We note that this solution does not depend strongly on the assumed value of $n_H$ in the range considered in this work (3 $\lesssim$ log$(n_H)$ $\lesssim$  8 [cm$^{-3}$]), as verified by \textsc{Cloudy} models.

While we have robust column density measurements from multiple troughs corresponding to \ion{Co}{II} and a single trough corresponding to \ion{Zn}{II}, we did not use them to constrain our photoionization solution because the ionization structure of the singly-ionized iron-peak elements is very similar to that of \ion{Fe}{II}. Their contours in the ($N_H$, $U_H$) phase space shown in Fig. \ref{fig:nvu} are thereofre parallel to that of \ion{Fe}{II}, and the exact location depends on their respective abundances. As Co and Zn are much less abundant than Fe, their ionization structure is more prone to be affected by a slight variation in their abundances as compared to Fe, and we therefore did not include them in our determination of the best-fit photoionization solution. 

\section{Number density of the outflow} \label{sec:den}

\subsection{Electron number density}    \label{sec:eden}

\begin{figure}[]
         \centering
         \includegraphics[width=1.0\linewidth]{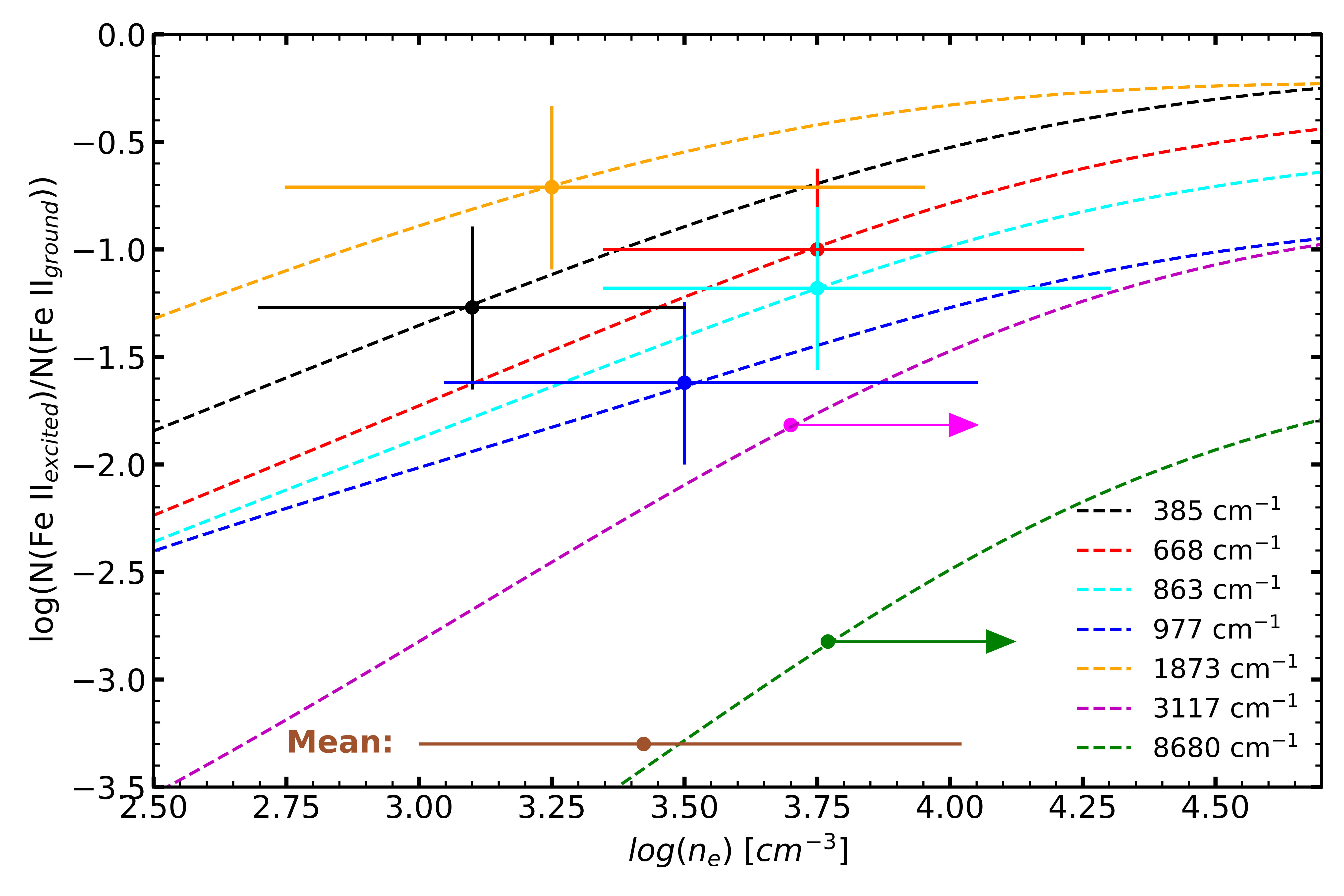}
         \caption{Theoretical population ratios for different \ion{Fe}{ii}* transitions for $T_e$ = 5000 K (determined from our \textsc{Cloudy} modeling; dashed curves). The solid crosses represent the observed column density ratios and their corresponding error (which includes the error determined in Sect. \ref{sec:osc}). The arrows represent the lower limits obtained from possibly saturated troughs corresponding to these energy levels. The mean $n_e$ along with its errors obtained from all the levels is shown in brown. }
         \label{fig:chianti}
\end{figure}

In the absorption system S2 of the outflow in J0932-0840, we detected multiple troughs arising from excited metastable levels of \ion{Fe}{II} (see Fig. \ref{fig:feiimodel}). Under the assumption of collisional excitation, the ratios of the column densities of these excited lines to the resonance line can be used as an indicator of the electron number density ($n_e$) of the outflow \citep[e.g.,][]{de2001keck, arav2018evidence}. To do this, we obtained the theoretical population ratios of the various excited states and the ground state as a function of $n_e$ using the \textsc{Chianti} atomic database \citep[vers. 9.0][]{dere1997chianti,dere2019chianti}. We then compared these predictions with our measurements of the column density ratios with the results shown in Fig. \ref{fig:chianti}. The various excited levels constrain log($n_e$) to be between 3.1 and 3.75 [cm$^{-3}$]. We determined its weighted mean using the method described by \cite{barlow2004asymmetric} and obtained the error using the procedure described in Appendix \ref{appendixA}. This resulted in log($n_e$) = $3.42^{+0.65}_{-0.46}$. 

In determining $n_e$, we used an effective electron temperature of $T_e$ = 5000 K based on the average \ion{Fe}{II} temperature determined from our photoionization modeling. This temperature is lower than the common assumption of $T_e$ $\approx$ 10,000 K for photoionized outflows. However, it does not significantly affect our solution for $n_e$ as its dependence on the temperature in this range is weak. This can be understood by noting that for a given energy level, the relative excited state population depends on the temperature through the Boltzmann factor $e^{-\Delta E/kT}$, where $\Delta E$ is the energy of the transition, and $k$ is the Boltzmann constant. For the \ion{Fe}{II*} levels used in determining $n_e$, $\Delta E$ $\leq$ 1873 cm$^{-1}$ and thus, $\Delta E/k$ $\lesssim$ 2700 K. Therefore, for higher temperatures, the contribution from the Boltzmann factor does not vary significantly, and the dependence of the relative excited state population on the temperature is thus weak. 

We also measured the column densities of troughs corresponding to two excited metastable levels of \ion{Ni}{II} (and a lower limit for the ground state). However, as \cite{dunn2010quasar} noted, \ion{Ni}{II} cannot be used as a reliable indicator of the electron number density of the outflow because as opposed to \ion{Fe}{ii}, \ion{Ni}{ii} is much more prone to fluorescence excitation by the UV continuum \citep{lucy1995fluorescent, bautista1996excitation}. Therefore, the population of the \ion{Ni}{II} excited levels is highly sensitive to the incident radiation, which makes it hard to model its excited-state population. We thus focused solely on \ion{Fe}{II} for our density diagnostics.

\subsection{Hydrogen number density} \label{sec:hden} 

We performed a similar analysis for the hydrogen number density ($n_H$) by modeling the photoionized gas using \textsc{Cloudy}, which offers two advantages over the \textsc{Chianti} analysis. First, \textsc{Cloudy} does not assume fixed values in $n_e$ and $T_e$ within the gas. For a given value of the gas density $n_H$ and a choice of elemental abundances, $n_e$, $T_e$, and the ion number densities are instead computed self-consistently given the depth-dependent thermal and ionization equilibrium solutions determined zone by zone throughout the cloud. Second, since an assumption about the outflow gas density $n_H$ is part of the ionization equilibrium solution, there is no need to guess its value based upon the value of $n_e$ determined from the level-population or excitation analysis.  The \textsc{Cloudy} models also include the contributions from fluorescence excitation (and other microphysics) in addition to collisional excitation, although this is not expected to significantly affect the relevant \ion{Fe}{II} level populations, as mentioned in the previous section.

\begin{figure}[]
         \centering
         \includegraphics[width=1.0\linewidth]{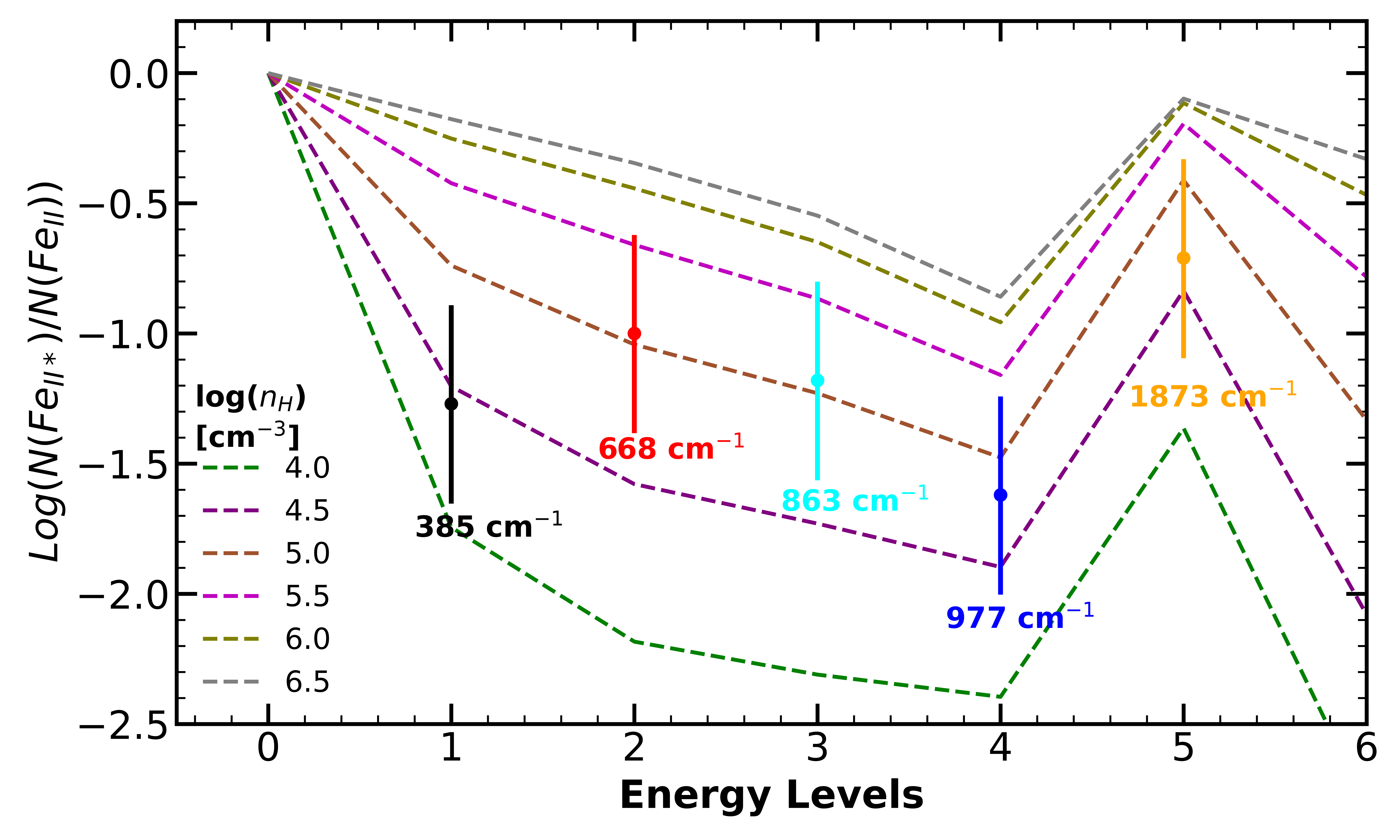}
         \caption{Excited-state to ground-state ratios for \ion{Fe}{II}. The dashed lines show the level population ratio predicted by \textsc{Cloudy} for different values of the hydrogen number density. The measurements are shown as colored circles and are accompanied by their errors.}
         \label{fig:cloudy_ratio}
\end{figure}

For our analysis, we ran models for the outflow with fixed $N_H$ and $U_H$ (obtained from our photoionization modeling in Sect. \ref{nvu}) for a range of $n_H$ ($10^{4.0} \leq n_H \textrm{ (cm}^{-3}) \leq 10^{6.5}$, in steps of 0.5 dex). These models predict the column densities for all the available levels of \ion{Fe}{II}, from which we obtained the column density ratios for the various levels to the ground state and compared them with the ratios obtained from the spectrum. The results of this analysis are shown in Fig. \ref{fig:cloudy_ratio}.

The different excited-state ratios were reproduced for hydrogen number densities between $4.5 \lesssim$ log($n_H$) $\lesssim 5.0$ [cm$^{-3}$]. To determine the best-fit value of $n_H$ from this analysis, we varied log($n_H$) between 4.0 and 5.5 [cm$^{-3}$] in steps of 0.1 to create a grid of models. We then compared the observed column density ratios of the excited states with the model predictions to perform a $\chi^2$ minimization. This resulted in a best-fit solution with log($n_H$) = $4.8^{}_{}$ [cm$^{-3}$]. 
We discuss the relation of the electron number density solution determined using \textsc{Chianti} (log($n_e$) = $3.42$ [cm$^{-3}$]) and the hydrogen number density solution determined from \textsc{Cloudy} (log($n_H$) = $4.8^{}_{}$ [cm$^{-3}$]) in the next section.

\subsection{Physical structure of the cloud} \label{sec:physical}

\begin{figure}
         \centering
         \includegraphics[width=1.00\linewidth]{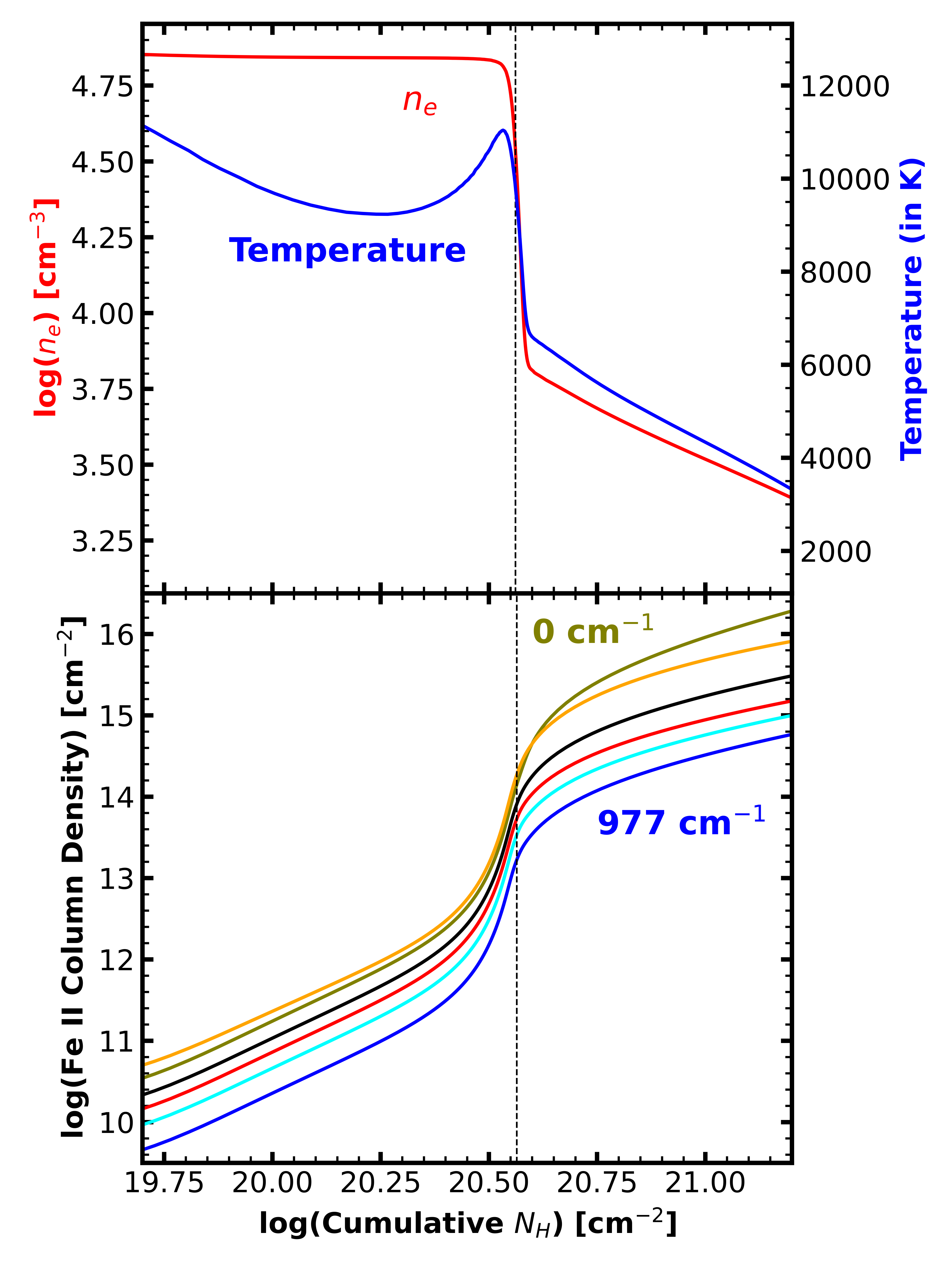}
         \caption{Physical structure of the photoionized cloud vs. the total hydrogen column density for log($n_H$) = 4.8 [cm$^{-3}$]. The top panels show the variation in the electron density and temperature within the cloud. The bottoms panels show the cumulative column density of each \ion{Fe}{II} level detected in the spectra. The dashed vertical black line marks the \ion{H}{I} ionization front.}
         \label{fig:physical}
\end{figure}

The formation of \ion{Fe}{II} and the excitation of various energy levels are better understood by a detailed examination of the physical structure of the photoionized cloud. As found through the photoionization modeling in Fig. \ref{fig:nvu}, the cloud is bound by a total hydrogen column density of log(N$_H$) = = $21.47$ [cm$^{-2}$]. Using log $U_H$ = $-2.4$ and the HE0238 SED, we employed \textsc{Cloudy} to provide a zone-wise description of the cloud for the best-fit hydrogen number density solution obtained in \ref{sec:hden} (log($n_H$) = $4.8^{}_{}$ [cm$^{-3}$]). Figure \ref{fig:physical} shows the obtained physical parameters ($T_e$ and $n_e$) and the column densities for the observed \ion{Fe}{II} levels as a function of the hydrogen column density within the cloud. We defined the hydrogen ionization front (marked by the dashed black line) as the point in the cloud at which half the hydrogen is ionized, and therefore, the density of fully ionized hydrogen (\ion{H}{ii}) and neutral hydrogen (\ion{H}{I}) is similar.

Figure \ref{fig:physical} (top panel, blue curve) shows that the temperature within the cloud drops suddenly across the \ion{H}{I} ionization front because most of the highly energetic photons are absorbed at the front. The electron number density also remains fairly constant before the \ion{H}{I} front. In this region, as most of the hydrogen is fully ionized, the electron number density is linearly related to the hydrogen number density with  $n_e \approx$ 1.2 $n_H$ \citep[approximation for fully ionized plasma;][]{osterbrock2006astrophysics}. However, across the \ion{H}{I} front, $n_e$ drops suddenly by around an order of magnitude, as shown in Fig. \ref{fig:physical} (top panel, red curve). This decrease is attributed to hydrogen becoming neutral and thus absorbing most of the free electrons. The bottom panel shows that the column densities of all observed levels of \ion{Fe}{ii} also vary rapidly near the front, and a sudden increase occurs just inside. The hydrogen ionization front is essential for the effective formation of \ion{Fe}{ii} as it acts as a shield against high-energy photons, thus preventing its further ionization. Therefore, most of the observed \ion{Fe}{II} comes from the region of the cloud beyond the \ion{H}{I} front. The varying electron number densities and temperatures in this region indicate that the conditions that determine the collisional excitation equilibrium are starkly different before and after the \ion{H}{I} ionization front. This is an important effect that shows that the majority of \ion{Fe}{ii} is subjected to a significantly reduced electron number density than the rest of the cloud, where it traces the constant hydrogen number density. Therefore, the approximation $n_e \approx$ 1.2 $n_H$ is not correct in FeLoBAL outflows, and using this approximation can lead to a severe underestimation for $n_H$. 

We estimated the effective $T_e$ and $n_e$ for \ion{Fe}{II} by averaging over all the zones while weighting their contribution based on the their relative \ion{Fe}{II} populations. This resulted in $T_{e_{eff}}$ = 5060 K and log($n_{e_{eff}}$) = 3.51 [cm$^{-3}$]. This agrees remarkably well with the solution obtained from \textsc{Chianti} in Sect. \ref{sec:eden} with log($n_e$) = $3.42^{+0.60}_{-0.42}$ [cm$^{-3}$]. This shows that the two solutions obtained independently for $n_e$ and $n_H$, respectively, are consistent with each other. We thus adopted log($n_H$) = $4.80^{+0.60}_{-0.52}$ [cm$^{-3}$] for the cloud. \\

We note that this is not the first time a difference between the two density solutions was deduced for a low-ionization quasar outflow. \cite{dunn2010quasar} found $n_H$ for their outflow to be $\sim$0.4 dex higher than $n_e$. Similarly, \cite{de2001keck} found that their models required $n_H$ to be $\sim$ 0.4/1.4 dex higher than $n_e$ depending on the choice of the abundance and depletion model. On the other hand, \cite{moe2009quasar} and \cite{bautista2010distance} reported that for their models, $n_e$ $\approx$ $n_H$. These different studies thus revealed that the low-ionization gas in these outflows is subject to a diverse range of physical conditions. 

\section{Distance and energetics} \label{energy}

\subsection{Distance to the outflow} \label{sec:distance}

After we obtained $U_H$ from photoionization modeling (Sect. \ref{nvu}) and $n_H$ (Sect. \ref{sec:den}), we inverted Eq. (\ref{uh}) to obtain the distance of the outflow from the emission source ($R$) as

\begin{equation} \label{eqn:distance}
    R = \sqrt{\frac{Q_H}{4\pi U_H n_H c}}.
\end{equation}

To obtain the rate of hydrogen ionizing photons emitted by the source, we scaled the HE0238 SED used in our modeling to match the estimated SDSS flux of J0932+0840 at an observed wavelength of  $\lambda \simeq$ 5680 $\AA$, with $F_{\lambda}$ = $7.35\times10^{-17}$ erg $s^{-1}$ cm$^{-2}$ $\AA^{-1}$. Integrating over the scaled SED for energy values above 1 Ryd. gives us $Q_H$ = $4.52_{-0.45}^{+0.45}$ $\times$ $10^{56}$ $s^{-1}$ and bolometric luminosity $L_{Bol}$ = $8.10_{-0.81}^{+0.81}$ $\times$ $10^{46}$ erg $s^{-1}$. This resulted in a location of the outflow that is $0.7_{-0.4}^{+0.9}$ kpc away from the central source. 

\subsection{Black hole mass and Eddington luminosity} \label{sec:bhmass}

As described in Sect. \ref{sec:redshift}, \ion{Mg}{II} 2799 $\AA$ is one of the most prominent emission features in the SDSS spectra. Based on the scaling relation derived in \cite{bahk2019calibrating}, we estimated the black hole mass ($M_{BH}$) from the \ion{Mg}{II} emission line as

\begin{multline} \label{MBH}
    \textrm{log}\left(\frac{M_{BH}}{M_{\odot}}\right) \approx (6.79 \pm 0.06) + \textrm{0.5 } (\textrm{log } L_{3000,44}) + \\ 2 \textrm{ log }(\textrm{ FWHM}_{\ion{Mg}{II}}) ,
\end{multline}

where $L_{3000,44}$ is the rest-frame luminosity $\lambda L_{\lambda}$ at 3000$\AA$ in units of $10^{44}$ erg $s^{-1}$, and the FWHM$_{\ion{Mg}{II}}$ is in units of 1000 km $s^{-1}$. Based on our modeling in Fig. \ref{fig:MgII}, FWHM$_{\ion{Mg}{II}}$ $\approx$ 3430 km $s^{-1}$, and from our continuum modeling of the SDSS spectrum at 3000 $\AA$, $\lambda L_{\lambda}$ = $2.86^{+0.43}_{-0.43}$ $\times$ $10^{46}$ erg $s^{-1}$. These values used with Eq. (\ref{MBH}) lead to log($\frac{M_{BH}}{M_{\odot}}$) = $9.09_{-0.1}^{+0.1}$, which results in an Eddington luminosity $L_{edd}$ = $1.54_{-0.30}^{+0.36}$ $\times$ $10^{47}$ erg $s^{-1}$. 
\subsection{Energetics}

In a simple geometry for the outflow in the form of a partial spherical shell that covers the fractional solid angle $\Omega$ and moves with a velocity $v$, the total mass is (see \cite{borguet201210} for a detailed discussion)
\begin{equation}
    M \simeq 4\pi\Omega R^2 N_H \mu m_p    ,
\end{equation}
where $R$ is the distance of the outflow from the source, $N_H$ is the total hydrogen column density of the outflow, $m_p$ is the proton mass, and $\mu$ = 1.4 is the atomic weight of the plasma per proton. Defining the dynamic timescale $t_{dyn}$ = $R/v$, where $v$ is the outflow velocity, we obtained the mass-flow rate $\dot{M}$ and the kinetic luminosity $\dot{E_k}$ ,

\begin{equation} \label{eqn:kinlum}
    \dot{M} = \frac{M}{t_{dyn}} = 4\pi\Omega R N_H \mu m_p v
\end{equation}
\begin{equation*}
    \dot{E_k} = \frac{1}{2} \dot{M} v^2 .
\end{equation*}

As $\Omega$ cannot be determined from the spectrum, the fraction of quasars that show the particular class of outflows has commonly been used as a substitute. Although FeLoBALs are extremely rare, as detailed in the introduction, we used $\Omega$ $\approx$ 0.2, based on the fraction of quasars that show \ion{C}{iv} BALs (which is seen in the SDSS spectra of J0932+0840). This is based on the argument by \cite{dunn2010quasar} based on the morphological indistinguishability of the \ion{C}{IV} BALs in FeLoBALs with \ion{C}{IV} BALs in the ubiquitous high-ionization BALQSOs. According to this, FeLoBALs fall under the class of normal $\Omega$ $\approx$ 0.2 outflows that are observed through a rare LOS.

This results in a mass-loss rate of $\dot{M}$ = $43^{+65}_{-26}$ $M_{\odot}$ yr$^{-1}$ and kinetic luminosity $\dot{E_k}$ = $0.7^{+1.1}_{-0.4}$ $\times$ $10^{43}$ erg $s^{-1}$, which is $0.5^{+0.7}_{-0.3}$ $\times$ $10^{-4}$ of the quasar $L_{Edd}$ and $0.9^{+1.3}_{-0.5}$ $\times$ $10^{-4}$ of its $L_{Bol}$ (see Table \ref{table:energetics} for a summary of the obtained parameters for the quasar and its outflow). \cite{hopkins2010quasar} showed that a kinetic luminosity of $\sim$ 5 $\times$ $10^{-3}$ of the quasar luminosity is required for an efficient AGN feedback. As the kinetic luminosity of this outflow component is much lower than this requirement, it is not expected to contribute significantly toward AGN feedback processes.

\begingroup
\setlength{\tabcolsep}{10pt}
\renewcommand{\arraystretch}{1.6}
\begin{table}
\caption{Physical parameters of quasar J0932+0840 and its outflow S2.}
\centering
\begin{tabular}{lc}
\hline
$v$ (km s$^{-1}$)         &-720\\
$\log{N_H}$ [cm$^{-2}$]         &$21.47^{+0.17}_{-0.16}$\\
$\log{U_H}$         &$-2.4^{+0.4}_{-0.5}$\\
$\log{n_H}$ [cm$^{-3}$]          &$4.80^{+0.60}_{-0.52}$\\
$R$ (kpc)          &$0.7_{-0.4}^{+0.9}$\\
$\dot{M}$ ($M_\odot$ yr$^{-1}$)          &$43^{+65}_{-26}$\\
$\log{\dot{E}_k}$ [erg s$^{-1}$]          &$42.85^{+0.40}_{-0.39}$\\
$L_{Edd}$ (erg s$^{-1}$)          &$1.54_{-0.30}^{+0.36}$ $\times$ $10^{47}$\\
$L_{Bol}$ (erg s$^{-1}$)          &$8.10_{-0.81}^{+0.81}$ $\times$ $10^{46}$\\
$\dot{E}_k/L_{Edd}$       &$0.5^{+0.7}_{-0.3}$ $\times$ $10^{-4}$\\
$\dot{E}_k/L_{Bol}$        &$0.9^{+1.3}_{-0.5}$ $\times$ $10^{-4}$\\
\hline
\end{tabular}
\label{table:energetics}
\end{table}
\endgroup

\section{Discussion} \label{discuss}

\subsection{Variability in the broad absorption lines} \label{sec:variability}

As mentioned in Sect. \ref{data}, the two different epochs of SDSS spectra were obtained $\sim$ 8 years apart, which spans roughly 8/(1+$z$) $\sim$ 2.4 yr in the rest frame of the quasar. We compared them over their common wavelength range (4000 - 9000 $\AA$ in the observed frame) and found no discernible change in the continuum as the flux level remained the same in regions without emission or absorption. The outflow identified in the UVES spectrum is also clearly detected in the SDSS spectra. However, due to its lower resolution, the three narrow low-velocity components ($v$ $\sim$ $-$500, $-$700, and $-$850 km s$^{-1}$) appear as a single component, whereas the broader high-velocity components ($v$ $\sim$ $-$1300, $-$4200, and $-$5000 km s$^{-1}$) are well resolved. The low-velocity component does not vary significantly over the two epochs for any of the observed ionic species. This is best confirmed by the nonblack troughs corresponding to the low-ionization species (e.g., \ion{Si}{II}, \ion{Fe}{II}, and \ion{Ni}{II}). For the high-ionization species (e.g, \ion{C}{IV}, \ion{Si}{IV}, and \ion{N}{V}), it is harder to determine any variation as their troughs are completely saturated, and thus, it cannot be completely ruled out.  

\begin{figure}[h]
         \centering
         \includegraphics[width=1.0\linewidth]{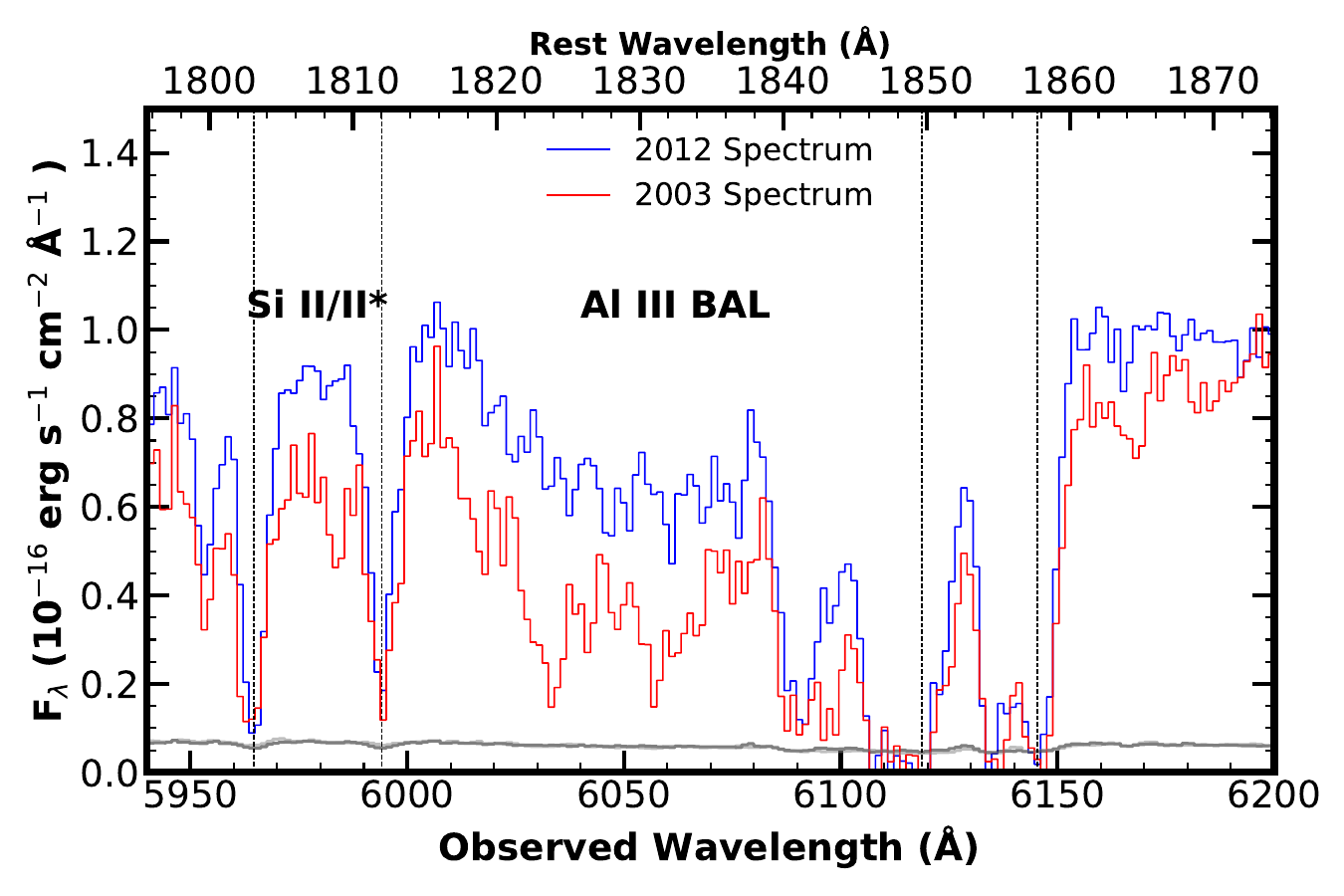}
         \caption{Spectral region around the \ion{Al}{III} BAL as observed in the 2003 (in red) and 2012 spectra (in blue). The troughs corresponding to our main outflow system at $v$ $\approx -$ 700 km $s^{-1}$ are marked with a dashed black line. We chose this spectral region as it shows the most noticeable difference in the high-velocity structure between the two epochs.}
         \label{fig:sdss_var}
\end{figure}

For the broad high-velocity components, however, a clear change is visible as most of the nonblack troughs become considerably shallower. Figure \ref{fig:sdss_var} shows this phenomenon for \ion{Al}{III}, which shows the most dramatic decrease among all observed species. The low-velocity components for \ion{Al}{III}, \ion{Si}{II}, and \ion{Si}{II}* (marked with vertical black lines) do not show any appreciable variation. However, the high-velocity \ion{Al}{III} component (between 6010 < $\lambda_{obs}$ < 6080 $\AA$) shows stronger absorption and a more pronounced kinematic structure for the 2003 spectrum than for the 2012 spectrum. Possible explanations for this variation in the high-velocity structure include (i) transverse motion of the outflowing gas across our line of sight or (ii) a change in the ionization state of the gas. In Sects. \ref{sec:transverse} and \ref{sec:ionization}, we consider these two scenarios. 

\subsubsection{Transverse motion} \label{sec:transverse}

In this case, variability in the troughs is assumed to be due to the transverse motion of the absorbing gas such that its coverage of the continuum source along our line of sight changes. With the help of some simplifying assumptions, we determined the transverse velocity of the variable component and an upper limit for its distance from the central source following previous studies (\citep[e.g.,][]{moe2009quasar, hall2011implications, yi2019variability}). Assuming that the black hole accretes close to the Eddington limit, we used its bolometric luminosity and black hole mass to estimate the radius of the accretion disk. Using Eq. (3.15) of \cite{peterson1997introduction} and the relation of the accretion disk size to black hole mass derived by \cite{morgan2010quasar}, we found the accretion disk radius to be about one light day.

The residual intensity of the \ion{Al}{III} component shown in Fig. \ref{fig:sdss_var} increases by $\sim$ 0.4 in 2.4 yr in the quasar rest frame. In the simplest scenario, this would imply that the transverse motion of the gas allowed it to cross roughly 40\% of the two-light-day-wide continuum source in the span of 2.4 yr. This results in a lower limit for the transverse velocity of $v_{tr} \gtrsim$ 275 km s$^{-1}$. Finally, when we assume that this tangential velocity does  not exceed the Keplerian orbital velocity solely due to the SMBH, we obtain an upper limit on the distance of the high-velocity component from the center of $R$ $<$ $G M_{BH}/v^2$ $\approx$ 70 pc.

However, at these large distances, the outflowing gas could be affected by the potential of the stars in the galactic bulge. We determined whether this contribution is significant by obtaining the radius of influence ($r_h$) of the SMBH. $r_h$ is defined as the radius at which the Keplerian rotation velocity equals the velocity dispersion of the galaxy, and it is thus a typical estimate of the radius within which the SMBH must dominate the gravitational potential \citep{peebles1972star}. Using Eqs. (17.1) and (17.6) of \textcolor{blue}{Bovy (in prep.)} \footnote{An online draft of the work is available at \url{https://galaxiesbook.org/}.}, we obtained $r_h$ $\approx$ 46 pc. As this is smaller than our upper limit on the distance (70 pc), we needed to include the contribution from the Galactic bulge. Within the sphere of influence, the total mass of the stars is roughly equal to that of the SMBH. We assumed that the density of the galactic bulge is roughly constant around the center (within the scale of a few hundred parsecs) and obtained the enclosed stellar mass as a function of the radius, with $M_{*}(<r)$ $\approx$ $M_{BH}$ ($r$/$r_h$)$^{3}$. Then, we solved for the Keplerian velocity profile as

\begin{equation}
    v(r) \approx \sqrt{G \left(\frac{M_{BH}+M_{*}(<r)}{r}\right)}  .
\end{equation}

This revealed that the Keplerian velocity decreases with distance from the center because the mass of stars is initially far lower than that of the SMBH. However, with increasing distance, continually more stellar mass is accumulated and eventually becomes the dominant factor. Therefore, the velocity eventually stops to decrease, with a minimum at $r \approx$ 0.8 $r_h$, beyond which it increases monotonically to the extent of the bulge. In our case, the minimum Keplerian velocity ($v_{min}$) $\approx$ 470 km s$^{-1}$ is higher than the lower limit on the transverse velocity obtained for the outflowing component. Therefore, we cannot obtain any estimate on the its distance based on the available information about the variability.

Finally, we note that while considering the case for transverse motion of the gas, we assumed that the variable broad structure (with $\Delta v$ $\gtrsim$ 3000 km $s^{-1}$) shown in Fig \ref{fig:sdss_var} is a single outflow component. If this broad feature were composed of different velocity components that could be at different radii from the central source, however, the observed variability would require their coordinated movement, which is unlikely. In this scenario, a change in the ionization state of the gas is a more viable explanation for the variability \citep[see Sect. 5.1 of ][for a detailed discussion]{capellupo2013variability}.

\subsubsection{Change in ionization state} \label{sec:ionization}

Because the ionization equilibrium of the outflow is dominated by photoionization, a change in the incident ionizing flux is expected to lead to a change in the ionization state of the absorbing gas, which would then be reflected in the corresponding troughs. In this section, we consider whether this mechanism might explain the nature of the absorption trough variability in the SDSS spectra of J0932+0840. Our model focused on three physical and observational constraints: 1. The plausibility of change in the ionizing flux, 2. a lack of variation in the \ion{Si}{II} 1808, 1817 \r{A} troughs for our main outflowing system, and 3. a significant change in the \ion{Al}{III} $\lambda \lambda$ 1855, 1863 doublet corresponding to a higher-velocity outflowing system. We discuss each of these in detail below. 

1. For J0932+0840, as the spectra do not cover the wavelength region responsible for the ionizing radiation ($\lambda_{rest}$ $\leq$ 912 $\AA$), we have no direct information about variation in the ionizing flux. In the spectral region covered by the two SDSS epochs (1200 $\lesssim$ $\lambda_{rest}$ $\lesssim$ 2700 $\AA$), the underlying continuum emission does not change appreciably, which indicates that no drastic changes occur in the shorter wavelength regime because they are usually assumed to be correlated. However, an exception to this correlation has been observed by \cite{arav2015anatomy}, and a variation in the ionizing flux therefore cannot be ruled out based on the far-UV flux alone. To study the effect of any possible variation in detail, we considered an example in which the ionizing flux increased by a factor of $\sim$ 2 between the two epochs. 

\begin{figure}[h]
         \centering
         \includegraphics[width=1.00\linewidth]{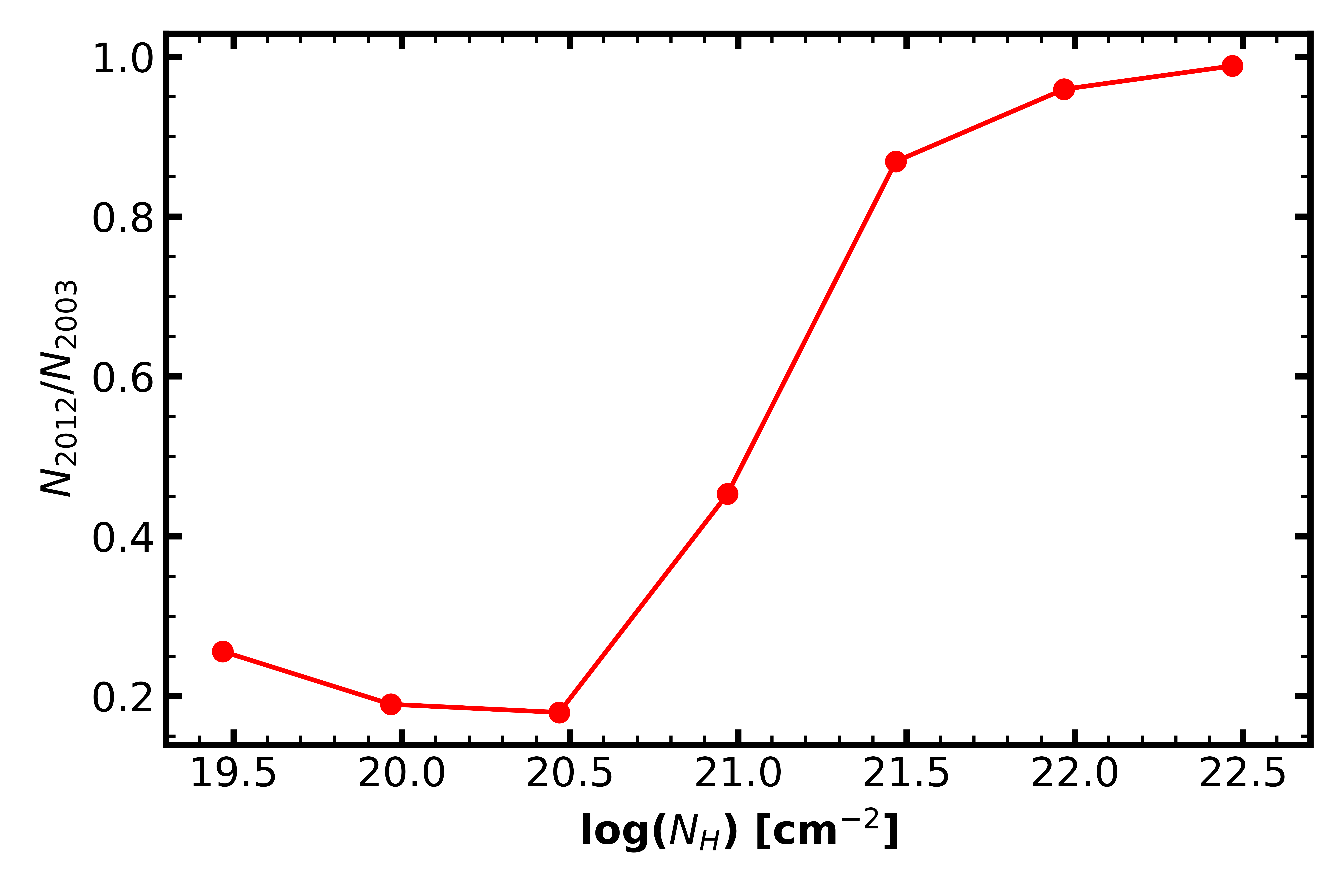}
         \caption{Change in the column density of \ion{Si}{II} due to the increase by a factor of 2 in the ionizing flux for clouds with a different total hydrogen column density ($N_H$). For clouds that did not form the \ion{H}{I} ionization front (log($N_H$) $\lesssim$ 20.5 [cm$^{-2}$]), the variation in column density is much higher than for the clouds that formed the ionization front.}
         \label{fig:siII_var}
\end{figure}

2. We first considered the main outflowing component at $v$ $\sim$ $-$ 700 km s$^{-1}$, for which no significant variation was detected in the nonblack troughs from the low-ionization species (e.g., see the \ion{Si}{II} troughs in Fig. \ref{fig:sdss_var}). To study the maximum variation in the ionization state of the gas as a result of the increase in the ionizing flux, we assumed that a new ionization equilibrium was established over a timescale shorter that the separation between the epochs. Equation \ref{uh} shows that the new ionization parameter ($U_{H_{2012}}$) should then be related to the old ionization parameter ($U_{H_{2003}}$) as $U_{H_{2012}}$/$U_{H_{2003}}$ $\sim$ 2, while other parameters such as $N_H$ and $n_H$ remain unchanged.  When we assume that the ionization parameter of this component during the 2003 epoch is similar to that obtained from the VLT/UVES spectrum, we have log($U_{H_{2003}}$) = $-$2.4 and log($U_{H_{2012}}$) = $-$2.1. We compared our photoionization models (with log($N_H$) = 21.5 [cm$^{-2}$]) of the outflowing gas in both these ionization states and found that the \ion{Si}{II} column density decreased by only $\sim$ 10\%  despite the 100\% increase in the ionizing flux. To better understand this phenomenon, we tracked the decrease in the \ion{Si}{II} column density as a function of $N_H$, and the results are shown in Fig. \ref{fig:siII_var}. For the same variation in the ionizing flux, the percentage change in \ion{Si}{II} column density can be different by up to an order of magnitude, depending on the total hydrogen column density of the cloud. 

This effect can be understood by noting that the formation of singly-ionized species such as \ion{Si}{II} strongly depends on the hydrogen ionization front (see Sect. \ref{sec:physical}), which forms around log($N_H$) $\sim$ 20.5 [cm$^{-2}$] in our case. Before the \ion{H}{I} front, only a small fraction of silicon is in the form of \ion{Si}{II}, and the majority is in more highly ionized forms (e.g., \ion{Si}{III/IV}). The exact amount of \ion{Si}{II} is thus determined by a delicate ionization balance with the higher ionized species, which strongly depends on the incident ionizing flux. Therefore, variations in this flux are reflected strongly in the \ion{Si}{II} column density before the \ion{H}{I} front. On the other hand, after the front, most of the silicon is found to be in the form of \ion{Si}{II} ($>$ 90\% for log($N_H$) $\geq$ 21.5 [cm$^{-2}$]). With the increase in the ionizing flux, the \ion{H}{I} front shifts to a higher $N_H$, but it still continues to shield the gas behind it from \ion{Si}{II} ionizing photons (i.e., photons with an energy greater than the ionization potential of \ion{Si}{II} $\sim$ 16.35 eV) and the ionization balance of silicon there continues to be dominated by \ion{Si}{II}. Therefore, the \ion{Si}{II} column density for the clouds with higher $N_H$ is much less sensitive to the change in the ionizing flux. This is also true for other singly ionized species (e.g., \ion{Fe}{II} and \ion{Ni}{II}), and thus, their column densities are not expected to vary significantly in the case of our main outflowing component either, which is consistent with our observations.

3. We considered the higher-velocity component at $v$ $\sim$ $-$5100 km s$^{-1}$, which shows troughs from multiple species such as \ion{Si}{II}, \ion{C}{II}, \ion{Al}{III}, \ion{Mg}{II}, \ion{Si}{IV}, and \ion{C}{IV}. The \ion{Al}{iii} $\lambda \lambda$ 1855, 1863 doublet for this component is shown in Fig. 9 between 6020-6070 \r{A}, where it shows considerable variation between the two epochs. There are also clear signs of variations for the unsaturated troughs of \ion{Si}{II} 1260 \r{A} and \ion{C}{II} 1335 \r{A}, but the troughs corresponding to \ion{Mg}{II}, \ion{Si}{IV}, and \ion{C}{IV} are saturated, and thus, their variation cannot be characterized properly. While the photoionization solution of this component is not well constrained, we considered whether this variation in the low-ionization components might be understood in terms of a change in the ionization state of the gas. \cite{krolik1996observable} provided an estimate for the recombination timescale ($t^*$) over which a sudden change in the incident ionizing flux would lead to a change in the ionic fraction of a species. For our main outflowing component, we find $t^*_{\ion{Al}{iii}}$ $\sim$ 1.24 yr. We lack a density diagnostic for the high-velocity component, and we thus cannot obtain its recombination timescale directly. However, we note that due to its much higher velocity, this component is expected to be closer to the central source and thus have a higher hydrogen and electron number density than our main outflowing component (from Eq. \ref{eqn:distance}). As $t^*$ is inversely proportional to $n_e$ for similar temperatures, we expect the recombination timescale of the high-velocity component to be significantly shorter than 1.24 yr. It would thus be much shorter that the 2.4 yr span between the two epochs in the quasar rest frame. Therefore, the new \ion{Al}{III} column density will agree well with that predicted by the time-independent photoionization equilibrium equations.

To study this variation quantitatively, we first obtained the column density of the \ion{Al}{III} 1863 \r{A} trough in the 2003 epoch using Eq. \ref{nion}. This was done by assuming an AOD model and integrating over the velocity range -5450 $\leq$ $v$ $\leq$ -4550 km s$^{-1}$, which resulted in log($N_{\ion{Al}{III}}$) = 14.81 [cm$^{-2}$]. We repeated this process (with the same velocity range) for the 2012 epoch and found log($N_{\ion{Al}{III}}$) = 14.49 [cm$^{-2}$]. If this decrease in \ion{Al}{iii} column density is due to a change in the ionization state, we should be able to find a constant value of $N_H$ and the two ionization parameters ($U_H$ and $U^{'}_H$) that can reproduce the observed \ion{Al}{III} column densities for the two epochs. We again assumed that the ionizing flux increased by a factor of 2 between the two epochs, and therefore, we have $U_H/U^{'}_H$ = 0.5. With this constraint in mind, we explored a $(N_H, U_H)$ phase-space plot (similar to Fig. \ref{fig:nvu}) for the measured \ion{Al}{III} column densities and obtained log($N_H$) = 21.1 [cm$^{-2}$], log($U_{H}$) = -2 and log($U^{'}_H$) = -1.7. This variation would also predict a decrease in the column densities of \ion{Si}{II} and \ion{C}{II}, which is also seen in the observations. On the other hand, the \ion{Si}{IV} and \ion{C}{IV} column densities would be expected to increase, but this cannot be confirmed as the corresponding troughs are saturated in both epochs. Therefore, our example shows that the observed variability pattern might be due to changes in the ionization state of the gas. 

\subsection{Previous high-resolution studies of FeLoBAL outflows} \label{previousFeLo}

In Sect. \ref{sec:den} we highlighted the importance of taking the physical structure of the cloud into account while estimating its $n_H$ from the $n_e$ determined from observed excited states of low-ionization species such as \ion{Fe}{II}. Depending on parameters such as the thickness of the cloud, $n_e$/$n_H$ can take values much lower than 1.2. The assumption that the plasma is highly ionized can then lead to an overestimation for the distance of the outflow and its energetics (see Eqs. \ref{eqn:distance} and \ref{eqn:kinlum}). \\
We found four such cases in the previous high-resolution studies of FeLoBAL outflows with a distance determination in which the effect of the physical structure of the cloud was not considered in their density diagnostics. For each of these objects, we modeled the outflowing cloud using the $N_H$, $U_H$ and the incident SED prescribed in the respective analyses. We then ran a grid of models with varying $n_H$ to determine the value for which the weighted $n_e$ matched the value determined using excited-state ratios observed in their outflow system. Using this new $n_H$, we then reevaluated $R$, $\dot{M}$, and $\dot{E} _{{k}}$ for the outflows. Table \ref{table:reevaluate} shows the result of this analysis. For three of these objects, the reevaluated distance determination does not vary significantly from the reported distance. However, for the outflow in J1321-0041, we find that $n_e/n_H$ $\lesssim$ $10^{-1}$, and therefore, the distance of the outflow from the central source was overestimated by $\sim$ 300\%.

\begin{table*}[ht!]
\setlength{\tabcolsep}{9pt}
\caption{Reevaluated density, distance, and energetics for some FeLoBAL outflows from the literature.}
\label{table1}
\centering
\renewcommand{\arraystretch}{1.4}
\begin{tabular}{l c c c c c c c c }
\hline
\textbf{Object}& log($N_H$) & log($U_H$) & log($n_H$) & $v$ &$R_{old}$ $^a$ &$R_{new}$ $^b$&\textbf{$\dot{M}$}&log $\dot{E} _{{k}}$   \\
 & [cm$^{-2}$]& &[cm$^{-3}$] & km s$^{-1}$ & kpc & kpc & $M_{\odot}$ yr$^{-1}$ & ergs s$^{-1}$  \\

\hline

J2357-0048$^{1}$ & $18.96^{+1.05}_{-0.64}$ & $-3.76^{+0.88}_{-0.77}$ & $4.32^{+0.2}_{-0.2}$ & $-$1600 & $10.90_{-7.1}^{+16.6}$  & $9.30_{-6.0}^{+14.0}$& $3800^{+96000}_{-3200}$ $^c$&$45.49^{+1.47}_{-0.85}$ $^c$ \\

J0242+0049$^{2}$ & $21.27^{+0.64}_{-0.58}$ & $-1.30^{+0.49}_{-0.48}$ & $0.21^{+0.21}_{-0.20}$ & $-$1800 & $67^{+55}_{-31}$  & $65^{+53}_{-30}$& $6300^{+12000}_{-3600}$&$45.81^{+0.47}_{-0.37}$ \\  

J1439-0106$^{3}$ & $20.99^{+0.61}_{-0.59}$ & $-1.99^{+0.60}_{-0.58}$ & $3.72^{+0.1}_{-0.1}$& $-$1550 & $2.60^{+2.50}_{-1.3}$ & $1.71^{+1.20}_{-1.07}$& $70^{+140}_{-50}$&$43.75^{+0.45}_{-0.50}$  \\    

J1321-0041$^{4}$ & $21.73_{-0.26}^{+0.39}$ & $-1.74^{+0.69}_{-0.27}$ & $4.56^{+0.26}_{-0.20}$ & $-$4100 & $2.50^{+1.00}_{-1.40}$ & $0.64^{+0.31}_{-0.37}$& $400^{+680}_{-260}$&$45.34^{+0.424}_{-0.451}$  \\   
\hline
\end{tabular}
\label{table:reevaluate}
\tablefoot{(a) Distance reported in the original work. (b) Reevaluated distance. (c) The energetics is for the higher-ionization phase of the outflow, whereas the density was obtained from the low-ionization phase. \\
\textit{References}: (1)\cite{byun_vltuves_2022-1}, (2) \cite{byun2022farthest}, (3) \cite{byun_vltuves_2022}, (4) \cite{byun_extreme_2024} }
\end{table*}

\section{Summary and conclusion} \label{summary}

We identified several outflow systems in the VLT/UVES spectrum of quasar SDSS J0932+0840, and analyzed the system with v $\sim$ -720 kms$^{-1}$ in detail. 

\begin{enumerate}
    \item In Sect. \ref{colden} we outlined our determination of the column densities of several ionic species. This enabled us to employ photoionization modeling (see Fig. \ref{fig:nvu}) for the outflow and obtain its total hydrogen column density ($\log{N_H}$ = $21.47^{+0.17}_{-0.16}$ [cm$^{-2}$]) and ionization parameter ($\log{U_H}$ = $-2.4^{+0.4}_{-0.5}$). 
    \item The detection of several ground- and excited-state transitions of \ion{Fe}{ii} allowed us to obtain the number density of the outflow. We modeled the electron number density ($n_e$) and hydrogen number density ($n_H$) independently and obtained log($n_e$) = $3.42^{+0.65}_{-0.46}$ [cm$^{-3}$] (Sect. \ref{sec:eden}) and log($n_H$) = $4.80$ [cm$^{-3}$] (Sect. \ref{sec:hden}).
    \item We examined the physical structure of the cloud in Sect. \ref{sec:physical} and showed that the formation of \ion{Fe}{II} is subject to a set of conditions different from most of the cloud, as it takes place beyond the hydrogen ionization front, where $n_e$ and temperature drop suddenly (see Fig. \ref{fig:physical}). For FeLoBAL outflows, the approximation of highly ionized plasma therefore does not hold as $n_e$ $\not\approx$ 1.2$n_H$. 
    \item We obtained the effective $n_e$ for our modeled cloud with log($n_H$) = $4.80$ [cm$^{-3}$] as log($n_{e_{eff}}$) = 3.51 [cm$^{-3}$]. This agrees with the measured value of log($n_e$) = $3.42^{+0.65}_{-0.46}$ [cm$^{-3}$], and we thus showed that our independent solutions for $n_e$ and $n_H$ are consistent with each other.
    \item Using our obtained solution for $n_H$, we estimated the distance of the outflow to be $0.7^{+0.9}_{-0.4}$ kpc from the central source. This led to $\dot{M}$ = $43^{+65}_{-26}$ $M_{\odot}$ yr$^{-1}$ and  $\dot{E_k}$ = $0.7^{+1.1}_{-0.4}$ $\times$ $10^{43}$ erg $s^{-1}$, with $\dot{E}_k/L_{Edd}$ = $0.5^{+0.7}_{-0.3}$ $\times$ $10^{-4}$. This outflow is thus not expected to contribute significantly to the AGN feedback. 
    \item We compared two epochs of SDSS spectra separated by $\sim$ 2.4 years in the quasar rest frame and explained the observed pattern of the absorption variation in terms of a change in the ionization state of the gas. 
 
\end{enumerate}

\begin{acknowledgements}
We acknowledge support from NSF grant AST 2106249, as well as NASA STScI grants AR-15786, AR-16600, AR-16601 and AR-17556.
\end{acknowledgements}


%
%

\bibliographystyle{aa}
\bibliography{J0932}

\onecolumn

\begin{appendix}

\section{Error estimation for mean $n_e$} \label{appendixA}

We observed various excited states of \ion{Fe}{II} in the system S2 of the outflow. Using the methodology described in Sect. \ref{sec:eden}, we obtained a corresponding electron number density ($n_e$) solution and its errors for each level as shown in Table \ref{table:appendix_ne}. To characterize the outflow with a single $n_e$, we obtain the weighted mean using the methodology described by \cite{barlow2004asymmetric}. To obtain an estimate for the error on the weighted mean, we use two different procedures, which are described below.

\begingroup
\setlength{\tabcolsep}{10pt}
\renewcommand{\arraystretch}{1.6}
\begin{table}[h]
\caption{$n_e$ obtained from \textsc{Chianti} for different \ion{Fe}{II} level ratios}
\centering
\begin{tabular}{|c|c|}
\hline
\textbf{Energy Level}     &     \textbf{log($n_e$)}  \\
(cm$^{-1}$) & [cm$^{-3}$] \\
\hline
 385      &$3.1^{+0.4}_{-0.35}$\\
 668      &$3.75^{+0.5}_{-0.4}$\\
 863       &$3.75^{+0.55}_{-0.4}$\\
 977     &$3.5^{+0.55}_{-0.45}$\\
 1873     &$3.25^{+0.7}_{-0.5}$\\
 \hline
\textbf{Weighted Mean} & 3.42 \\
\hline
\end{tabular}
\label{table:appendix_ne}
\end{table}
\endgroup

\subsection{Propagating errors associated with each measurement}

One way to estimate error for the mean is to obtain the mean of the weighted errors added in quadrature. If the positive/negative error for the $i^{th}$ measurement of $x$ ($x_i$) is denoted as $(\delta x_i)^{\pm}$, then the positive/negative error for the mean, $(\delta \left<x\right>)^{\pm}$, can be obtained as: 

\begin{equation} \label{eqn:A1}
    (\delta \left<x\right>)^{\pm} = \frac{\sum\sqrt{(w_i \cdot (\delta x_i)^{\pm})^2}}{\sum w_i}
\end{equation}
where $w_i$ is the weight associated with each measurement as described in \cite{barlow2004asymmetric}. With $x$ = log($n_e$), we obtain $(\delta \left<\textrm{log}(n_e)\right>)^{+}$ = 0.24 and $(\delta \left<\textrm{log}(n_e)\right>)^{-}$ = 0.14, and therefore we have $\left<\textrm{log}(n_e)\right>$ = $3.42_{-0.19}^{+0.24}$. While this is a robust method to take into account the errors associated with each measurement, it does not include information about the scatter in the values between measurements obtained from the different levels. \\

\subsection{Standard deviation including the measurement errors}

To obtain errors on the mean in a way which includes the spread of the measurements around it, and also their own individual errors, we define:

\begin{equation} \label{eqn:A2}
    (\delta \left<x\right>)^{\pm} = \sqrt{\frac{\sum ( \left<x\right> - (x_i \pm (\delta x_i)^{\pm} ))^2}{N}}
\end{equation}
where $N$ is the total number of measurements and $\left<x\right>$ is their mean. With $x$ = log($n_e$), this definition leads to $(\delta \left<\textrm{log}(n_e)\right>)^{+}$ = 0.65 and $(\delta \left<\textrm{log}(n_e)\right>)^{-}$ = $0.46$, and therefore we have $\left<\textrm{log}(n_e)\right>$ = $3.42_{-0.46}^{+0.65}$. We adopt this value for our analysis and represent it graphically in Fig. \ref{fig:chianti}. It shows that given the scatter between the different measurements, Eq. (\ref{eqn:A2}) provides a much better representation of the error compared to the smaller values obtained from Eq. (\ref{eqn:A1}).

\end{appendix}

\end{document}